\documentclass[]{article}

\input glyphtounicode
\usepackage[T1]{fontenc}
\usepackage[utf8]{inputenc}

\usepackage{ifdraft}

\usepackage[english]{babel}

\usepackage{mathtools}
\usepackage{enumitem}
\usepackage[amsmath,thmmarks,hyperref]{ntheorem}

\usepackage{xcolor}
\definecolor{hypercolor}{rgb}{0,0.2,0.7}
\usepackage[breaklinks,final]{hyperref}

\usepackage{footmisc}

\newif\iffancyfont%
\fancyfontfalse%

\IfFileExists{MinionPro.sty}{\IfFileExists{MyriadPro.sty}{\fancyfonttrue}{}}{}

\iffancyfont%
  \usepackage[lf,medfamily,italicgreek]{MinionPro}
  \usepackage[lf,medfamily,onlytext,scale=0.96]{MyriadPro}

  \renewcommand{\dotsb}{{\mathinner{\cdotp\cdotp\cdotp}}}
  \renewcommand{\dotsm}{{\mathinner{\cdotp\cdotp\cdotp}}}

  \DeclareRobustCommand{\defn}{\mathrel{{\vdotdot}{\equal}}}
  \DeclareRobustCommand{\nfed}{\mathrel{{\equal}{\vdotdot}}}

  \DeclareMathSymbol{\widehatsym}{\mathord}{largesymbols}{'302}

  \usepackage[toc,eqno,bib]{tabfigures}
\else%
  \usepackage[charter,greekuppercase=italicized,cal=cmcal]{mathdesign}

  \newcommand{\defn}{\coloneqq}
  \newcommand{\nfed}{\eqqcolon}

  \DeclareMathSymbol{\widehatsym}{\mathord}{largesymbols}{"62}

  \let\updelta\deltaup%
  \let\uppi\piup%
  \let\upDelta\Delta%

  \let\righthalfcup\lrcorner

  \makeatletter
  \newsavebox{\@brx}
  \newcommand{\llangle}[1][]{\savebox{\@brx}{\(\m@th{#1\langle}\)}%
    \mathopen{\copy\@brx\mkern2mu\kern-0.9\wd\@brx\usebox{\@brx}}}
  \newcommand{\rrangle}[1][]{\savebox{\@brx}{\(\m@th{#1\rangle}\)}%
    \mathclose{\copy\@brx\mkern2mu\kern-0.9\wd\@brx\usebox{\@brx}}}
  \makeatother
\fi%

\DeclareFontFamily{U}{BOONDOX-calo}{\skewchar\font=45}
\DeclareFontShape{U}{BOONDOX-calo}{m}{n}{
  <-> s*[0.96] BOONDOX-r-calo}{}
\DeclareFontShape{U}{BOONDOX-calo}{b}{n}{
  <-> s*[0.96] BOONDOX-b-calo}{}
\DeclareSymbolFont{bcal}{U}{BOONDOX-calo}{m}{n}
\SetSymbolFont{bcal}{bold}{U}{BOONDOX-calo}{b}{n}
\DeclareSymbolFontAlphabet{\mathcal}{bcal}
\DeclareFontFamily{U}{BOONDOX-scr}{\skewchar\font=45}
\DeclareFontShape{U}{BOONDOX-scr}{m}{n}{
  <-> s*[0.96] BOONDOX-r-cal}{}
\DeclareFontShape{U}{BOONDOX-scr}{b}{n}{
  <-> s*[0.96] BOONDOX-b-cal}{}
\DeclareSymbolFont{bscr}{U}{BOONDOX-scr}{m}{n}
\SetSymbolFont{bscr}{bold}{U}{BOONDOX-scr}{b}{n}
\DeclareSymbolFontAlphabet{\mathscr}{bscr}

\IfFileExists{inconsolata.sty}{\usepackage{inconsolata}}{}

\ifdraft{
  \let\what\hat%
  \let\wwhat\hat%
}{
  \usepackage{scalerel}
  \usepackage{accents}
  \newcommand\lowerwidehatsym{%
    \text{\smash{\kern-.1ex\raisebox{-1.3ex}{%
      $\widehatsym$}}}}
  \newcommand\what[1]{%
    \mathchoice%
      {\accentset{\displaystyle\ignoremathstyle\hstretch{0.5}\lowerwidehatsym}{#1}}%
      {\accentset{\textstyle\ignoremathstyle\hstretch{0.5}\lowerwidehatsym}{#1}}%
      {\accentset{\scriptstyle\ignoremathstyle\hstretch{0.5}\lowerwidehatsym}{#1}}%
      {\accentset{\scriptscriptstyle\ignoremathstyle\hstretch{0.5}\lowerwidehatsym}{#1}}%
  }
  \newcommand\wwhat[1]{%
    \mathchoice%
      {\accentset{\displaystyle\ignoremathstyle\hstretch{0.7}\lowerwidehatsym}{#1}}%
      {\accentset{\textstyle\ignoremathstyle\hstretch{0.7}\lowerwidehatsym}{#1}}%
      {\accentset{\scriptstyle\ignoremathstyle\hstretch{0.7}\lowerwidehatsym}{#1}}%
      {\accentset{\scriptscriptstyle\ignoremathstyle\hstretch{0.7}\lowerwidehatsym}{#1}}%
  }
}

\usepackage{microtype}

\usepackage[sf,bf,small]{titlesec}
\usepackage[labelfont={sf,bf},labelsep=period]{caption}

\setlist[1]{labelindent=\parindent}
\setlist[description]{font=\sffamily\bfseries,align=right,labelsep=1em}

\numberwithin{equation}{section}

\makeatletter
\newcounter{and}
\newdimen{\instindent}
\newcommand{\institute}[1]{\newcommand{\@institute}{#1}}
\newcommand{\inst}[1]{\unskip\smash{$^#1$}\setcounter{and}{1}\ignorespaces}
\newcommand{\email}[1]{\href{mailto:#1}{#1}}
\renewcommand{\maketitle}{
  {
    \raggedright%
    \Large%
    \noindent%
    \bfseries%
    \sffamily%
    \@title%
    \par
  }

  \vspace{1.5\baselineskip}

  {
    \raggedright%
    \renewcommand{\and}{\unskip, \ignorespaces}%
    \noindent\ignorespaces\@author\par
  }

  \vspace{0.5\baselineskip}

  {
    \small%
    \parindent=0pt%
    \parskip=0pt%
    \setcounter{and}{1}%
    \renewcommand{\and}{%
      \par\stepcounter{and}%
      \hangindent\instindent%
      \noindent%
      \hbox to \instindent{\hss\smash{$^{\theand}$\enspace}}\ignorespaces%
    }%
    \setbox0=\vbox{\@institute}%
    \ifnum\value{and}>9\relax\setbox0=\hbox{$^{88}$\enspace}%
    \else\setbox0=\hbox{$^{8}$\enspace}\fi%
    \instindent=\wd0\relax%
    \ifnum\value{and}=1\relax%
    \else%
      \setcounter{and}{1}%
      \hangindent\instindent%
      \noindent%
      \hbox to \instindent{\hss\smash{$^{\theand}$}\enspace}\ignorespaces%
    \fi%
    \ignorespaces%
    \@institute\par
  }
}
\makeatother

\renewenvironment{abstract}{
  \addvspace{1.5\baselineskip}%
  \topsep=0pt\partopsep=0pt%
  \trivlist\item[\hspace{\labelsep}\bfseries\sffamily Abstract.]
}{}

\newenvironment{acknowledgments}{
  \addvspace{1.5\baselineskip}%
  \topsep=0pt\partopsep=0pt%
  \trivlist\item[\hspace{\labelsep}\bfseries\sffamily Acknowledgments.]
}{}

\usepackage[obeyDraft]{todonotes}

\ifdraft{

  \usepackage[notref,notcite]{showkeys}

  \usepackage{filemod}
  \usepackage{fancyhdr}
  \pagestyle{fancy}
  \fancyhf{}
  
  \fancyhead[C]{Draft version. Last modified: \filemodprint{\jobname}}
  \fancyfoot[C]{\thepage}
}{}

\input{macros.def}

\theoremnumbering{arabic}
\theoremstyle{plain}
\theorembodyfont{\upshape}
\theoremheaderfont{\normalfont\sffamily\bfseries}
\theoremseparator{.}
\newtheorem{definition}{Definition}[section]

\title{Electromagnetic Potential in Pre-Metric Electrodynamics: Causal Structure, Propagators and Quantization}

\author{
  Christian Pfeifer\inst{1}${}^,$\inst{2}
  \and
  Daniel Siemssen\inst{3}
}

\institute{
  Institute for Theoretical Physics, Leibniz Universität Hannover, Appelstrasse 2, 30167 Hannover, Germany.\\
  E-mail: \email{christian.pfeifer@itp.uni-hannover.de}
  \and
  Center of Applied Space Technology and Microgravity (ZARM), University of Bremen, Am Fallturm, 28359 Bremen, Germany.
  \and
  Department of Mathematical Methods in Physics, Faculty of Physics, University of Warsaw, Pasteura 5, 02-093, Warszawa, Poland.\\
  E-mail: \email{daniel.siemssen@fuw.edu.pl}

}

\hypersetup{
  unicode,
  colorlinks,
  linkcolor=hypercolor,
  urlcolor=hypercolor,
  citecolor=hypercolor,
  bookmarksnumbered,
  bookmarksdepth=2,
  pdfauthor={Christian Pfeifer, Daniel Siemssen},
  pdftitle={Electromagnetic Potential in Pre-Metric Electrodynamics: Causal Structure, Propagators and Quantization}
}

\begin{document}

\maketitle

\begin{abstract}
  An axiomatic approach to electrodynamics reveals that Maxwell electrodynamics is just one instance of a variety of theories for which the name electrodynamics is justified.
  They all have in common that their fundamental input are Maxwell's equations $\dif F = 0$ (or $F = \dif A$) and $\dif H = J$ and a constitutive law $H = \conlaw F$ which relates the field strength two-form~$F$ and the excitation two-form~$H$.
  A local and linear constitutive law defines what is called local and linear pre-metric electrodynamics whose best known application is the effective description of electrodynamics inside media including, \eg, birefringence.
  We analyze the classical theory of the electromagnetic potential~$A$ before we use methods familiar from mathematical quantum field theory in curved spacetimes to quantize it in a locally covariant way.
  Our analysis of the classical theory contains the derivation of retarded and advanced propagators, the analysis of the causal structure on the basis of the constitutive law (instead of a metric) and a discussion of the classical phase space.
  This classical analysis sets the stage for the construction of the quantum field algebra and quantum states.
  Here one sees, among other things, that a microlocal spectrum condition can be formulated in this more general setting.
\end{abstract}


\section{Introduction}

One of the oldest physical field theories is Maxwell's description of the electromagnetic field and charged currents.
A more accurate description of the field and the currents requires the quantization of the theory and its embedding into the standard model of particle physics.
Today it serves as prototype theory of a gauge field theories.

Taking an axiomatic approach to classical electrodynamics one sees that Maxwell electrodynamics is only one subclass of a larger set of gauge theories which are all justified to be called electrodynamics.
Among them the theory which we will analyze and quantize in this article: local and linear pre-metric electrodynamics.

Assuming only\footnote{At least on a contractible manifold.} conservation of charge and magnetic flux, the most general formulation of electrodynamics is~\cite{Hehl}
\begin{subequations}\label{eq:premetric}\begin{align}
  \dif F &= 0, \\
  \dif H &= J.
\end{align}\end{subequations}
Above, the \emph{electromagnetic field strength} $F$ is an untwisted $2$-form, the \emph{electromagnetic excitation}~$H$ is a twisted $2$-form and the \emph{electric current} $J$ is a twisted $3$-form such that $\dif J = 0$ (it is closed, \viz, electric current is conserved).

These equations are underdetermined and need to be supplemented by a relation $H = H(F)$ between $H$ and $F$; this relation contains the physics of electrodynamics.
The most general local and linear theory of electrodynamics is obtained from a linear dependence of~$H$ on~$F$.
That is, one specifies a \emph{constitutive law}
\begin{equation}\label{eq:conlaw}
  H(F) = \conlaw F
\end{equation}
by defining a invertible, pointwise,\footnote{By this we mean that $\conlaw$ descends from a map $\conlaw_x : TM^{0,2}_x(M) \to TM^{0,2}_x(M)$ at each $x \in M$ to assure locality.} linear map $\conlaw : \Omega^2(M) \to \Omega^2(M)$ which maps untwisted $2$-forms into twisted $2$-forms.
In the course of this article we will restrict ourselves first to non-dispersive constitutive laws and later to those which lead to a causal behaviour.
For a large part of this article we also restrict to constant constitutive laws but the generalization to the non-constant case is more a technicality than a fundamental hurdle; most of our results should generalize immediately.

In \emph{Maxwell electrodynamics} the map~$\conlaw$ is given by the Hodge dual with respect to a Lorentzian metric, typically the Minkowski metric.
However, in general the constitutive law $\conlaw$ need not arise from a Lorentzian metric.
Therefore the classical field theory with field equations~\eqref{eq:premetric} and~\eqref{eq:conlaw} is called \emph{local and linear pre-metric electrodynamics}. For brevity we will call it simply \emph{pre-metric electrodynamics} in what follows.

These equations can be used to give a geometric description of electromagnetic fields in media~\cite{Hehl2005249, Hehl20081141} including polarization dependent refraction of light in crystals (birefringence)~\cite{Perlick}.
Moreover, pre-metric electrodynamics is a suitable generalization to describe electromagnetic fields in the presence of gravity induced vacuum polarization~\cite{PhysRevD.22.343} and can be used as a theory of electrodynamics in so-called area metric spacetimes~\cite{Punzi:2007di}.
Certainly this pre-metric electrodynamics does not cover the description of electrodynamics in all types of media.
A large variety of media are know in which non-local or non-linear constitutive laws are needed to describe the behavior of the electromagnetic field correctly.
The discussion of those theories is beyond the scope of this article.

Besides the presentation of the quantization of pre-metric electrodynamics one objective of this article is to be accessible for readers with a background in pre-metric electrodynamics as well as readers with a background in algebraic quantum field theory and quantum  field theory on curved spacetime.
For this reason we are sometimes more verbose than absolutely necessary to improve readability for both targeted audiences.

A previous approach to quantize pre-metric electrodynamics using the canonical quantization method is discussed in~\cite{Rivera:2011rx}.
The aim of this work is to extend and complement~\cite{Rivera:2011rx} in some aspects from a different point of view.
In Sect.~\ref{sec:quant} we use the formalism of algebraic quantum field theory to quantize pre-metric electrodynamics.
To be more precise, we will follow roughly the approach of~\cite{dimock:1980}, which proved to be very useful in the context of quantum field theory on curved spacetimes.
We believe algebraic QFT to be the appropriate choice in the absence of a preferred vacuum state (as on curved spacetimes and more general geometries), see \eg\ the discussion in~\cite[Chap.~4]{wald:1994}.
In such a situation algebraic QFT gives us a mathematically rigorous toolbox to analyze quantum fields in a qualitative way.
Of course, for concrete calculations it is typically necessary to choose a state whence one can return to a Hilbert space setting via the Gelfand–Naimark–Segal (GNS) theorem.

Throughout this article we develop the classical theory of pre-metric electrodynamics Sect.~\ref{sec:classical} in view of what we need to construct its quantum version in Sect.~\ref{sec:quant}.

After an introduction of the field equation and our basic assumptions on the constitutive law in Sect.~\ref{sub:field_eq}, we analyze and invert its principal symbol in Sect.~\ref{sub:quasiinv}.
Due to the gauge-freedom present in the theory, the resulting object is only an inverse up to a gauge choice but we can classify this freedom precisely.
Moreover, we find that the principal symbol can only be inverted when the so-called Fresnel polynomial is non-zero.
If the Fresnel polynomial is hyperbolic, we show in Sect.~\ref{sub:causality} how it can be used to endow the manifold with a causal structure.
Then we introduce (partial) gauge-fixing operators in Sect.~\ref{sub:gauge}.
These are used in Sects.~\ref{sub:inv} and~\ref{sub:causp} to introduce advanced and retarded inverses (Green's solutions) and the Pauli--Jordan propagator.
Using the Pauli--Jordan propagator, we construct in Sect.~\ref{sub:sympl} spaces of solutions to the homogeneous field equations and equip them with a symplectic structure; these are the phase spaces to be quantized in Sect.~\ref{sec:quant}.
Finally, we will discuss the energy-momentum associated to the electric field in Sect.~\ref{sub:energydensity}.
This will culminate in the definition of a positive `energy product' on the space of solutions if the constitutive law satisfies certain conditions.
During our analysis of the classical theory we emphasize in particular which are the important properties of the classical theory that are required to construct the corresponding quantum field theory.

In Sect.~\ref{sub:alg} we will discuss the algebraic quantization of the classical phase space introduced in Sect.~\ref{sec:classical} and in particular in Sect.~\ref{sub:sympl}.
For this purpose we will introduce the field algebra of the electromagnetic potential.
The next step is the introduction of quantum states in Sect.~\ref{sub:states} and the discussion of their properties.
From the algebraic point of view, states are certain functionals on the field algebra, \ie, they are used to evaluate configurations of quantum fields.
However, not all states can be considered physical.
Therefore we will introduce the concept of normal ordering and the microlocal spectrum condition in Sect.~\ref{sub:norder}.
To make the relatively abstract content of Sect.~\ref{sec:quant} slightly more concrete, we complement it by the construction of a ground state in Sect.~\ref{sub:concrete_state}.
This construction is based on the energy product introduced in Sect.~\ref{sub:energydensity} and follows closely the construction of states for quantum fields on static spacetimes.

With the quantization of pre-metric electrodynamics we explicitly demonstrate that even field theories which do not rely on a spacetime metric but instead on a different geometric background, here defined by the constitutive law, can be quantized in a locally covariant fashion.

As discussed above, it turned out that pre-metric electrodynamics is a fruitful theory to describe physical effects.
We expect its quantized version to be useful when the interactions with the medium can be understood in an averaged classical sense but the quantum nature of light is important.

Let us close this introduction by specifying some conventions and notation:
If not otherwise specified, we consider complex-valued functions (and more generally sections) and function spaces.
Often we use index notation with Latin indices $a, b, \dotsc$ running from $0$ to $3$; the Einstein summation convention is always assumed.
We emphasize that due to the absence of a metric indices can generally not be raised or lowered.

\section{Classical field theory}\label{sec:classical}

As stated above, our main goal is the quantization of pre-metric electrodynamics using methods of algebraic quantum field theory.
For this purpose we first of all need a comprehensive understanding of the classical field theory.
After a discussion of the field equation governing pre-metric electrodynamics via the electromagnetic potential, we will analyze the corresponding Cauchy problem, construct the advanced and retarded Green's operators and derive the Pauli-Jordan propagator.
The latter enables us to covariantly introduce the symplectic phase space of the theory.
As in Maxwell electrodynamics, this analysis is closely intertwined with the gauge freedom of the theory.
An auxiliary result of our derivation of the fundamental solutions is a natural gauge condition which can be considered to be a generalization of the Lorenz gauge.

\subsection{Field equation}\label{sub:field_eq}

The field equations of pre-metric electrodynamics are derived from general electrodynamics~\eqref{eq:premetric} by inserting the linear constitutive law~\eqref{eq:conlaw}
\begin{subequations}\begin{align}
  \dif F &= 0, \\
  \dif \conlaw F &= J. \label{eq:genlin3form}
\end{align}\end{subequations}
Assuming a contractible manifold $M$, we have $F = \dif A$ so that the equations of general electrodynamics reduce to
\begin{equation}\label{eq:potential}
  P A \defn \dif \conlaw \dif A = J
\end{equation}
in terms of the \emph{electromagnetic (co)vector potential}~$A$, which is a $1$-form.
As in Maxwell electrodynamics, we find that two potentials $A, A'$ differing by a $1$-form $\dif \lambda$ solve the same equation; $A$ and~$A'$ are called gauge equivalent and are related by the gauge transformation $A \to A' = A + \dif \lambda$.

With help of the totally antisymmetric Levi-Civita epsilon symbol normalized such that $\epsilon_{0123}=1$ with respect to some positively oriented basis, the local and linear constitutive law can be expressed in local coordinates as
\begin{equation}\label{eq:conlawcoord}
  H_{ab} = (\conlaw F)_{ab}
  \nfed \frac{1}{2} \kappa_{ab}{}^{cd} F_{cd}
  \nfed \frac{1}{4} \varepsilon_{abcd} \chi^{cdef} F_{ef},
\end{equation}
where the relation between $\kappa$ and $\chi$ is given by
\begin{equation}
  \kappa_{ab}{}^{ef} = \frac{1}{2} \varepsilon_{abcd} \chi^{cdef}
  \quad\text{and}\quad
  \chi^{abef} = \frac{1}{2} \varepsilon^{abcd} \kappa_{cd}{}^{ef}.\label{eq:condens}
\end{equation}
We will always assume that $\kappa$ and $\chi$ depend smoothly on the base point of the manifold $M$.
By definition, $\kappa$ and $\chi$ have the symmetries
\begin{equation*}
  \kappa_{ab}{}^{cd} = \kappa_{[ab]}{}^{[cd]},
  \qquad
  \chi^{abcd} = \chi^{[ab][cd]}
\end{equation*}
and $\chi$ is a tensor density of weight $1$.
Moreover, we shall always assume the additional symmetry
\begin{equation}\label{eq:addsymm}
  \chi^{abcd} = \chi^{cdab}
  \quad\Leftrightarrow\quad
  F \wedge \conlaw F' = F'\! \wedge \conlaw F
\end{equation}
\viz, the constitutive law defines at each point a symmetric bilinear form on $2$-forms.
A more general linear constitutive law would lead to dissipative forces~\cite[Chap.~D.1.5]{Hehl} but both the classical and quantum description of a dissipative system is beyond the scope of this work.
Nevertheless, some results derived in this work hold independently of the assumption~\eqref{eq:addsymm}.

Using the coordinate expression~\eqref{eq:conlawcoord} of the constitutive law and $J = \frac{1}{3!} J_{abc}\, \dx^a\! \wedge \dx^b\! \wedge \dx^c$ in the field equation~\eqref{eq:potential}, we obtain
\begin{equation}\label{eq:genlincoord}
  \frac{1}{2}\partial_{[a} \big( \kappa_{bc]}{}^{de} \partial_d A_e \big)
  = \frac{1}{4} \partial_{[a} \big( \varepsilon_{bc]de} \chi^{defg} \partial_f A_g \big)
  = \frac{1}{3!} J_{abc}.
\end{equation}
Here the relation between the gauge freedom of the theory and the use of conserved currents becomes nicely visible in the symmetry properties of the constitutive density $\chi$.
The antisymmetry in the first index pair implements that $J$ is a conserved current, while the antisymmetry in the second index pair causes gauge invariance of the field equation under the transformation $A \to A' = A + \dif \lambda$.

It is easy to see that, applying twice Stokes' theorem and the symmetry~\eqref{eq:addsymm},
\begin{equation}\label{eq:P_selfadj}
  \int_M A \wedge P B
  = \int_M A \wedge \dif \conlaw \dif B
  = \int_M B \wedge \dif \conlaw \dif A
  = \int_M B \wedge P A
\end{equation}
for $1$-forms $A, B$ if $\supp A \cap \supp B$ is compact and $(\supp A \cup \supp B) \cap \partial M = \emptyset$.
In other words, \emph{$\int \cdot \wedge (P\,\cdot\,)$ is a symmetric bilinear form on the compactly supported $1$-forms}.
This should be understood as the generalization of the statement ``$P$ is formally self-adjoint'' to the case studied here.

Suppose that the constitutive law is given by the Hodge operator~$\hodge$ determined by a Lorentzian metric~$g$ (\eg, in the absence of gravity by the Minkowski metric), namely $H = \hodge F$.
Then the field equations become the well-known standard Maxwell equations
\begin{align}
  \dif F &= 0, \nonumber\\
  \dif \hodge F &= J, \nonumber
\shortintertext{or, equivalently,}
  \dif \hodge \dif A &= J. \label{eq:maxwell-potential}
\end{align}
To make the relation of~\eqref{eq:potential} and~\eqref{eq:maxwell-potential} manifest, we note that~\eqref{eq:potential} for
\begin{equation}\label{eq:constmax}
  \kappa_{cd}{}^{ab}
  = \abs{g}^\frac12 \varepsilon_{cdef} g^{ea} g^{fb}
  \quad\Leftrightarrow\quad
  \chi^{abcd} = 2 \abs{g}^\frac12 g^{a[c} g^{d]b}
\end{equation}
becomes
\begin{equation*}
  \dif \conlaw \dif A
  = \dif \big( \abs{g}^\frac12 \varepsilon_{cdef} g^{ea} g^{fb} \partial_a A_b\, \dx^c\! \wedge \dx^d \big)
  = J,
\end{equation*}
which is obviously identical to $\dif \hodge \dif A = J$.
Throughout this article we will call electrodynamics as described by~\eqref{eq:maxwell-potential} \emph{Maxwell electrodynamics} as opposed to \emph{pre-metric electrodynamics} described by~\eqref{eq:potential} with a generic linear and local constitutive law.

A more complex physical example of pre-metric electrodynamics are uniaxial crystals.
These are linear permeable media whose dielectricity is characterized by a spacelike vector field X and whose magnetic permeability is trivial, as measured by an observer given by a timelike vector field $U$ that is normalized $g(U,U) = -1$.
The constitutive density for uniaxial crystals becomes
\begin{equation}\label{eq:condensuni}
  \chi^{abcd}=\abs{g}^\frac12(2g^{c[a}g^{b]d}+4 X^{[a}U^{b]}X^{[d}U^{c]}).
\end{equation}
A derivation of this constitutive density can be found in App.~\ref{app:Uniaxial}.

These are just two examples of physical theories which are contained in the framework of pre-metric electrodynamics.
We now proceed towards solving the field equations by studying them in Fourier space.

\subsection{Inverting the principal symbol}\label{sub:quasiinv}

The partial differential operator of the field equations $P$ maps $1$-forms to closed $3$-forms $P : \Omega^1(M) \to \Omega^3_\dif(M)$ (the subscript $\dif$ indicates that the $\Omega^3_\dif(M)$ is the space of \emph{closed} $3$-forms).
Using~\eqref{eq:genlincoord}, we see that in a local coordinate basis it takes the form
\begin{equation*}
  P = P(x, -\im\partial) = \frac{1}{2}\big( \kappa_{ab}{}^{ed}(x) \partial_{c} \partial_e + (\partial_{a} \kappa_{bc}{}^{ed})(x) \partial_e \big)\, \dx^a\! \wedge \dx^b\! \wedge \dx^c \otimes \partial_d,
\end{equation*}
\ie,
\begin{equation*}
  P_{abc}{}^d = P_{abc}{}^d(x, -\im\partial) = \frac{3!}{2} \big( \kappa_{[ab}{}^{ed}(x) \partial_{c]} \partial_e + (\partial_{[a} \kappa_{bc]}{}^{ed})(x) \partial_e \big).
\end{equation*}
The \emph{principal symbol} of a partial differential operator $P$ is the leading order term in the polynomial $P(x, k)$ labelled by covectors $k$.
It is given by
\begin{equation}\label{eq:principal}
  M(x, k) \defn \conlaw (k \wedge \cdot\, ) \wedge k= \frac{1}{2}\kappa_{ab}{}^{ed}(x) k_c k_e\, \dx^a\! \wedge \dx^b\! \wedge \dx^c \otimes \partial_d,
\end{equation}
\ie, $M_{bcd}{}^a(x, k) \defn \frac{3!}{2} \kappa_{[bc}{}^{ea}(x) k_{d]} k_e$.
For constant $\kappa_{ab}{}^{cd}$, the principal symbol at $x$ can also be understood as the Fourier space representation of the field equations~\eqref{eq:potential}: $M(k) \what{A}(k) = \what{J}(k)$.
Note that the principal symbol $M(x, k)$ is covariantly defined as function on the cotangent bundle with values in the $(1,3)$-tensor fields on spacetime, \ie, in the vector fields with values in the $3$-forms.
In the following we will often suppress the explicit $x$ and $k$ dependence of $M$ and derived objects.

The principal symbol is at the core of the analysis of a partial differential equation.
On the one hand it determines the propagation of singularities of the solutions which we discuss briefly in App.~\ref{app:propsing} and on the other hand its inverse, which we will construct here, is the fundamental ingredient in the construction of an inverse of the field equations.
To obtain the desired inverse of the symbol $M$ it turns out to be most practical to introduce an equivalent symbol $\mathcal{M}^{ab}$ via
\begin{align}\label{eq:MtoM}
  \mathcal{M}^{ab} \defn \frac{1}{3!} \varepsilon^{acde} M_{cde}{}^b
  \quad\Leftrightarrow\quad
  M_{abc}{}^d = \varepsilon_{eabc} \mathcal{M}^{ed}.
\end{align}
This definition yields
\begin{equation*}
  \mathcal{M}^{ab} = \mathcal{M}^{(ab)} = \chi^{acbd} k_c k_d.
\end{equation*}

We seek for a quasi-inverse of the principal symbol since $\mathcal{M}(x,k)$ is degenerate by the symmetries of $\chi^{abcd}$
\begin{equation}\label{eq:Mdegeneracy}
  \mathcal{M}^{ab}(x,k) k_a = 0 = \mathcal{M}^{ab}(x,k) k_b
\end{equation}
and so an inverse does not exist.
This degeneracy reflects the gauge freedom, which in Fourier space reads $\what{A} \to \what{A} + \lambda k$, and the conservation of electric current $k \wedge \what{J} = 0$.
As already observed in~\cite{Itin:2009aa}, \eqref{eq:Mdegeneracy} reflects the deep interrelation between gauge freedom and the conservation of electric current -- they are dual to each other.
It may be seen as a consequence of an elementary theorem from linear algebra, according to which column and row rank of a matrix must agree.
Even though $\mathcal{M}(x,k)$ is not invertible we will see in the remainder of this section that it is possible to obtain an object which comes as close to a true inverse as is necessary to construct the inverse fo the field equations in Sect.~\ref{sub:inv}.
We will call this object quasi-inverse.

As a consequence of~\eqref{eq:Mdegeneracy}, the maximum rank of $\mathcal{M}$ is $3$.
Assuming for now that the rank of $\mathcal{M}$ is indeed $3$, there exists a three-dimensional subspace $V \subset (\CC^4)^* = T^*_xM \otimes \CC$ of the complexified%
\footnote{%
  Later we will consider covectors that have a ``small'' complex part thus we already allow for such complex covectors here.
}
cotangent space such that the restriction of $\mathcal{M}$ to this subspace, denoted by $\mathbf{M}$, is non-degenerate, \viz, the determinant of the restricted matrix $\mathbf{M} \in V \otimes V$ is non-zero.
The inverse of $\mathbf{M}$ is readily calculated by taking the quotient of its adjugate by its determinant:
\begin{equation}\label{eq:bfM_inverse}
  \mathbf{M}^{-1} = \frac{\adj(\mathbf{M})}{\det(\mathbf{M})}.
\end{equation}
In the following we will identify the spaces $V$, produce a covariant expression of $\mathbf{M}^{-1}$ and project it into the whole space $(\CC^4)^*$.

The process of restricting $\mathcal{M}$ to $\mathbf{M}$ corresponds to removing one row and one column from~$\mathcal{M}$ in a certain basis.
Recall that, up to transpositions, the (second) adjugate of a matrix~$\mathcal{M}$ is the matrix of determinants of the first (second) minors of $\mathcal{M}$.
Therefore the determinant $\det(\mathbf{M})$ can be identified with a component of the adjugate $\adj(\mathcal{M})$ in some basis and the adjugate $\adj(\mathbf{M})$ can be identified with components of the second adjugate $\adj_2(\mathcal{M})$ in the same basis.

Let $\kappa \in \CC^4$ be a dual vector to $k \in (\CC^4)^*$, \ie, it satisfies $\kappa^a k_a = 1$.
Note that every such dual vector specifies a three-dimensional space $V = \ker \kappa$.
An explicitly covariant formulation of the last sentence in the paragraph above is
\begin{subequations}\label{eq:bfM_M_relation}\begin{align}
  \det(\mathbf{M}) &= \adj(\mathcal{M})_{ab} \kappa^a \kappa^b, \\
  \adj(\mathbf{M})_{ab} &= \adj_2(\mathcal{M})_{cdef} (\delta^c_a - \kappa^c k_a) (\delta^d_b - \kappa^d k_a) \kappa^e \kappa^f \label{eq:bfM_M_relation2}.
\end{align}\end{subequations}
The factors $\pi^a_b \defn \delta^a_b  - \kappa^a k_b$ in~\eqref{eq:bfM_M_relation2} are projectors from $T^*_x M$ into $V$; it is clear that
\begin{equation}\label{eq:M_projector}
  \mathcal{M}^{ac} \pi^b_c = \mathcal{M}^{ab} = \mathcal{M}^{cb} \pi^a_c
\end{equation}
as a consequence of~\eqref{eq:Mdegeneracy}.
Using the second adjoint for the derivation of the quasi-inverse follows ideas by Itin~\cite{Itin:2007aa} and, more recently,~\cite{Itin:2015tdi}.

The same author showed~\cite{Itin:2007aa,Itin:2009aa,Itin:2015tdi} that the adjugate of $M$ is given by
\begin{equation}\label{eq:fresnel_poly}
  \adj(\mathcal{M})_{ab}(x, k) = \mathcal{G}(x, k) k_a k_b,
\end{equation}
where $\mathcal{G}$ is the so-called \emph{Fresnel polynomial}.
This equation is a consequence of~\eqref{eq:Mdegeneracy}, which implies that $\adj(\mathcal{M})$ has either rank $1$ or rank $0$, and the identity
\begin{equation*}
  \adj(\mathcal{M})_{ab} \mathcal{M}^{bc} = \mathcal{M}^{ab} \adj(\mathcal{M})_{bc} = 0.
\end{equation*}

Taking the definition of the adjugate and~\eqref{eq:fresnel_poly}, we calculate (independently of $\kappa$)
\begin{align}
  \mathcal{G}(x, k)
  &\defn \mathcal{G}(x; k,k,k,k) \defn \mathcal{G}^{abcd}(x) k_a k_b k_c k_d \defn \adj(\mathcal{M})_{ab} \kappa^a \kappa^b \\
  &\mathrel{\phantom{\defn}\mathllap{=}} \frac{1}{3!} \varepsilon_{a a_1 a_2 a_3} \varepsilon_{b b_1 b_2 b_3} \chi^{a_1 c_1 b_1 d_1} \chi^{a_2 c_2 b_2 d_2} \chi^{a_3 c_3 b_3 d_3} k_{c_1} k_{d_1} k_{c_2} k_{d_2} k_{c_3} k_{d_3} \kappa^a \kappa^b \\
  &\mathrel{\phantom{\defn}\mathllap{=}} \frac{1}{3!} \frac{1}{2} \varepsilon_{c_1 a_1 a_2 a_3} \varepsilon_{b b_1 b_2 b_3} \chi^{a_1 c_1 b_1 d_1} \chi^{a_2 c_2 b_2 d_2} \chi^{a_3 c_3 b_3 d_3} k_{d_1} k_{c_2} k_{d_2} k_{c_3} k_{d_3} \kappa^b \\
  &\mathrel{\phantom{\defn}\mathllap{=}} \frac{1}{4!} \varepsilon_{c_1 a_1 a_2 a_3} \varepsilon_{d_3 b_1 b_2 b_3} \chi^{a_1 c_1 b_1 d_1} \chi^{a_2 c_2 b_2 d_2} \chi^{a_3 c_3 b_3 d_3} k_{d_1} k_{c_2} k_{d_2} k_{c_3},
\end{align}
where $\mathcal{G}^{abcd}$ is called the \emph{Fresnel tensor density} and $\mathcal{G}(x, k)$ is a scalar density of weight $1$.
The equality in the third line follows from
\begin{equation}\label{eq:epsilon_identity}
  \varepsilon_{abcd} \chi^{abfg} u^c v^d = \varepsilon_{abcd} \big( \chi^{aefg} k_e \kappa^b + \chi^{ebfg} k_e \kappa^a \big) u^c v^d = 2 \varepsilon_{abcd} \chi^{aefg} \kappa^b k_e u^c v^d
\end{equation}
for arbitrary vectors $u, v$ in the kernel of $k$; an analogous identity shows the equality in the fourth line.
An alternative, elegant, index-free representation of the Fresnel polynomial which uses dyadics can be found in \cite{Lindell}.

In Maxwell electrodynamics $\chi^{acbd} =  2\abs{g}^{1/2} g^{a[b} g^{d]c}$ the Fresnel density decomposes into a symmetrized tensor square of the Lorentzian metric $g$
\begin{align}\label{eq:maxwell_fresnel}
  \mathcal{G}(k) = \abs{g}^\frac12 g^{-1}(k,k)^2
\end{align}
For linear permeable media $\chi^{acbd} = \abs{g}^{1/2} (2 g^{b[a} g^{c]d} + 4 X^{[a} U^{c]} X^{[d} U^{b]})$ we obtain a bi-metric Fresnel density which allows for birefringence since it vanishes if either of its two distinct metric factors vanishes
\begin{align}\label{eq:fresneluni}
  \mathcal{G}(k) = \abs{g}^\frac12 g^{-1}(k,k) \big( g^{-1}(k,k) - U(k)^2 g(X,X) + X(k)^2 \big);
\end{align}
for its derivation we refer to App.~\ref{app:Uniaxial}.

The Fresnel tensor density and the Fresnel polynomial play a central role in the analysis of the partial differential equation~\eqref{eq:potential} -- essentially, the Fresnel tensor defines the underlying causal structure of pre-metric electrodynamics~\cite{Schuller:2009hn}, while requiring hyperbolicity of the Fresnel polynomial guarantees that the initial value problem of the field equations is well posed~\cite{Hoermander2}, independent of the existence of a Lorentzian metric.
We will discuss the connection between causality and the properties of the Fresnel polynomial in more detail in Sect.~\ref{sub:causality}.

Of particular importance will be the Fresnel equation
\begin{equation}\label{eq:fresnel}
  \mathcal{G}(x, k) = 0,
\end{equation}
which determines the so-called characteristic wave covectors $k$ along which the singularities of the solutions to the field equations propagate, see App.~\ref{app:propsing}.
In pre-metric electrodynamics these are interpreted as light rays representing the geometrical optics limit of the theory.
We can already easily see that the Fresnel equation is satisfied if and only if the rank of $\mathcal{M}^{ab}(x, k)$ is less than~$3$.

Assume for now that $\mathcal{G}$ and $k$ are such that the Fresnel equation~\eqref{eq:fresnel} is \emph{not} satisfied.
That is, we assume that k is not a characteristic wave covector.
Only in this case $\det{\mathbf{M}}$ is non-vanishing and the inverse of $\mathbf{M}$ as well as the qusi-inverse of $\mathcal{M}$ can be constructed.
Note, however, that the existence of non-trivial solutions to~\eqref{eq:fresnel} is of great importance.
These non-trivial characteristic wave covectors determine the support properties of the Green’s functions which we construct in \ref{sub:inv}.

As can be seen from~\eqref{eq:bfM_inverse} and~\eqref{eq:bfM_M_relation}, the next step to construct the quasi-inverse of~$\mathcal{M}$ is to derive its second adjugate $\adj_2(\mathcal{M})$ twice contracted with the vector $\kappa$ and twice contracted with the projector $\pi^a_b = \delta^a_b  - \kappa^a k_b$:
\begin{align*}
  \adj_2(\mathcal{M})_{cdef} \pi^c_a \pi^d_b \kappa^e \kappa^f
  &= \frac{1}{2} \varepsilon_{d e a_1 a_2} \varepsilon_{c f b_1 b_2} \chi^{a_1 c_1 b_1 d_1} \chi^{a_2 c_2 b_2 d_2} k_{c_1} k_{d_1} k_{c_2} k_{d_2} \pi^c_a \pi^d_b \kappa^e \kappa^f \\
  &= \frac{1}{4} \varepsilon_{d c_1 a_1 a_2} \varepsilon_{c f b_1 b_2} \chi^{a_1 c_1 b_1 d_1} \chi^{a_2 c_2 b_2 d_2} k_{d_1} k_{c_2} k_{d_2} \pi^c_a \pi^d_b \kappa^f \\
  &= \frac{1}{8} \varepsilon_{d c_1 a_1 a_2} \varepsilon_{c d_2 b_1 b_2} \chi^{a_1 c_1 b_1 d_1} \chi^{a_2 c_2 b_2 d_2} k_{d_1} k_{c_2} \pi^c_a \pi^d_b,
\end{align*}
where we applied twice~\eqref{eq:epsilon_identity} as in the derivation of the Fresnel polynomial and used that $\pi^a_b v^b \in \ker k$ for all vectors $v$.
This leads us to the definition of the symmetric tensor
\begin{equation}
  \mathcal{Q}_{ab}(x, k) \defn \frac{1}{8} \varepsilon_{b c_1 a_1 a_2} \varepsilon_{a d_2 b_1 b_2} \chi^{a_1 c_1 b_1 d_1} \chi^{a_2 c_2 b_2 d_2} k_{d_1} k_{c_2}
\end{equation}
so that $\adj_2(M)_{cdef} \pi^c_a \pi^d_b \kappa^e \kappa^f = \mathcal{Q}_{cd} \pi^c_a \pi^d_b$.

To get an idea how this object looks like, we have a quick look at the special cases of Maxwell electrodynamics $\chi^{acbd} = 2 \abs{g}^{1/2} g^{a[b} g^{d]c}$
\begin{align}\label{eq:maxwell_Q}
  \mathcal{Q}_{ab} = g_{ab} g^{-1}(k,k) - k_a k_b,
\end{align}
and the uniaxial crystal $\chi^{acbd} = \abs{g}^{1/2} (2 g^{b[a} g^{c]d} + 4 X^{[a} U^{c]} X^{[d} U^{b]})$
\begin{align}\label{eq:uniaxial_Q}
  \mathcal{Q}_{ab}
  &= g_{ab} \big(g^{-1}(k,k) - X(k)^2 - U(k)^2 g(X,X) \big) + \big( X(k) U_a - U(k) X_a \big) \big( X(k) U_b - U(k) X_b \big) \nonumber\\
  &\mathrel{\phantom{=}}{} -k_{(a} \big( X_{b)} X(k) - U_{b)} U(k) g(X,X) \big) - k_a k_b.
\end{align}
The latter is derived in App.~\ref{app:Uniaxial}.

Although it is already clear from~\eqref{eq:bfM_inverse}, it is useful to see explicitly that $\mathcal{G}^{-1} \mathcal{Q}_{cd} \pi^c_a \pi^d_b$ is an inverse of $\mathcal{M}^{ab}$ restricted to $V = \ker \kappa$.
The second adjugate satisfies the identities
\begin{align*}
  \mathcal{M}^{ae} \adj_2(\mathcal{M})_{ebcd} &= \delta^a_b \adj(\mathcal{M})_{cd} - \delta^a_d \adj(\mathcal{M})_{cb} = \mathcal{G} (\delta^a_b k_c k_d - \delta^a_d k_b k_c), \\
  \mathcal{M}^{ea} \adj_2(\mathcal{M})_{becd} &= \delta^a_b \adj(\mathcal{M})_{cd} - \delta^a_c \adj(\mathcal{M})_{bd} = \mathcal{G} (\delta^a_b k_c k_d - \delta^a_c k_b k_d)
\end{align*}
and, using~\eqref{eq:M_projector}, we thus find
\begin{equation}\label{eq:MQ}
  \mathcal{M}^{ac} \mathcal{Q}_{cd} \pi^d_b = \mathcal{M}^{ca} \mathcal{Q}_{dc} \pi^d_b = \mathcal{G} \pi^a_b.
\end{equation}
Thus, restricted to the subspace $V$, \emph{$\mathcal{M}$ can be inverted}.
Since $V$ is the kernel of $\kappa$ and thus $\pi^a_b(x, k)$ is a projector from $T^*_x M$ into $V$, this shows that
\begin{equation}
  \wwhat{\mathcal{E}}_{ab}(x, k)
  \defn \mathcal{G}^{-1} \mathcal{Q}_{cd} \pi^c_a \pi^d_b
  = \mathcal{G}^{-1} \mathcal{Q}_{cd} (\delta^c_a - \kappa^c k_a) (\delta^d_b - \kappa^d k_b).
\end{equation}
is the inverse of $\mathcal{M}$ if we restrict to contractions with covectors in $V$, \ie, $\mathcal{M}^{ca}\wwhat{\mathcal{E}}_{ab}=\pi^c_b$.

We stress that $\wwhat{\mathcal{E}}$ depends on the choice of the vector $\kappa$.
If $\kappa'$ is another vector dual to $k$ and $\pi'$ the corresponding projector,
then
\begin{equation*}
  \wwhat{\mathcal{E}}'_{ab}(x, k) = \mathcal{G}^{-1} \mathcal{Q}_{cd} \pi'{}^c_a \pi'{}^d_b
\end{equation*}
is another inverse of $M$ and it is related to $\wwhat{\mathcal{E}}$ via
\begin{equation}\label{eq:different_kappa}
  \wwhat{\mathcal{E}}'_{ab} = \wwhat{\mathcal{E}}_{cd} \pi'{}^c_a \pi'{}^d_b
  \quad\text{or, equivalently,}\quad
  \wwhat{\mathcal{E}}_{ab} = \wwhat{\mathcal{E}}'_{cd} \pi^c_a \pi^d_b
\end{equation}
because $\pi^a_c \pi'{}^c_b = \pi'{}^a_b$ and $\pi'{}^a_c \pi^c_b = \pi^a_b$.
Expanding the product in~\eqref{eq:different_kappa} and rearranging terms, we obtain
\begin{equation*}
  \wwhat{\mathcal{E}}_{ab} = \wwhat{\mathcal{E}}'_{ab} - \big( \wwhat{\mathcal{E}}'_{cb} - \tfrac{1}{2} \wwhat{\mathcal{E}}'_{cd} \kappa^d k_b \big) \kappa^c k_a - \big( \wwhat{\mathcal{E}}'_{ad} - \tfrac{1}{2} \wwhat{\mathcal{E}}'_{cd} \kappa^c k_a \big) \kappa^d k_b,
\end{equation*}
which simplifies to
\begin{equation}\label{eq:quasiinv_trafo}
  \wwhat{\mathcal{E}}_{ab} = \wwhat{\mathcal{E}}'_{ab} - m_b k_a - m_a k_b
  \quad\text{with}\quad
  m_a = \wwhat{\mathcal{E}}'_{ab} \kappa^b - \wwhat{\mathcal{E}}'_{bc} \kappa^b \kappa^c k_a / 2,
\end{equation}
where we used that $\mathcal{Q}_{(ab)} = \mathcal{Q}_{ab}$ due to~\eqref{eq:addsymm}.
Observe that directly from~\eqref{eq:quasiinv_trafo} follows $\mathcal{M}^{ab} m_b = \kappa^a - \kappa'^a$ so that, given two of $m, \kappa, \kappa'$, we can recover the third.
These gauge transformations of the quasi-inverse fit in the structure of gauge transformations Itin found for the photon propagator in linear response media \cite{Itin:2015tdi}.
Looking at the Maxwell case~\eqref{eq:maxwell_Q}, we notice that~\eqref{eq:quasiinv_trafo} is exactly the photon propagator transformation Eq.~(76.5) of~\cite{lifshitz:1982}.
As a further relation with the treatment of quantum electrodynamics, we remark that $\pi$ corresponds to a polarization sum over three polarizations vectors.

Having discussed these similarities to Maxwell electrodynamics we construct the quasi-inverse to the original principal symbol $M$.
In light of the relation~\eqref{eq:MtoM} we define
\begin{align}
  Q_a{}^{bcd} &= \frac{1}{3!} \varepsilon^{ebcd} \mathcal{Q}_{ae} \label{eq:defQ3form}
\shortintertext{and}
  \wwhat{E}_a{}^{bcd} &= \frac{1}{3!} \varepsilon^{ebcd} \wwhat{\mathcal{E}}_{ae} = \frac{1}{3!} \varepsilon^{ebcd} \mathcal{G}^{-1} \mathcal{Q}_{fg} \pi^f_a \pi^g_e. \label{eq:qinv3form}
\end{align}
One might still wonder about the role of the projectors $\pi$ in the equations above.
As we will see later in even more detail, they are related to fixing the gauge freedom in electrodynamics.
For now let us just remark that for a $3$-form $\what{J} = \frac{1}{3!} \what{J}_{abc}\, \dx^a\! \wedge \dx^b\! \wedge \dx^c$ and a $1$-form $\what{A} = \what{A}_a\, \dx^a$ we have
\begin{subequations}\begin{align}
  M_{abc}{}^d \wwhat{E}_d{}^{efg} \what{J}_{efg}
  &= \frac{1}{3!} \varepsilon_{pabc} \mathcal{M}^{pd} \varepsilon^{qefg} \wwhat{\mathcal{E}}_{dq} \what{J}_{efg}
  = \frac{1}{3!} \varepsilon_{pabc} \varepsilon^{qefg} \pi^p_q \what{J}_{efg}
  = \what{J}_{abc} \label{eq:fourier_inv1} \\
\shortintertext{and}
  \wwhat{E}_a{}^{bcd} M_{bcd}{}^e \what{A}_e
  &= \wwhat{\mathcal{E}}_{ac} \mathcal{M}^{cb} \what{A}_b
  = \pi^b_a \what{A}_b
  = \what{A}_a + \lambda k_a, \label{eq:fourier_inv2}
\end{align}\end{subequations}
if $k \wedge \what{J} = 0$ and $\lambda = \kappa^a \what{A}_a$, which is again the conservation of the electric current $3$-form and a gauge transformation (actually a gauge fixing).
Observe that a solution of $M_{abc}{}^d \what{A}_d = \what{J}_{abc}$ generated by $\wwhat{E}$ from the conserved current $\what{J}$ via $\what{A}_a = \wwhat{E}_a{}^{bcd} \what{J}_{bcd}$ satisfies the gauge condition $\kappa^a \what{A}_a = \lambda = 0$ since $\kappa^a \pi^b_a = 0$.

Before we study the gauge properties of the theory in more detail, we make a short detour to introduce notions of causality in the context of pre-metric electrodynamics in terms of the Fresnel polynomial.

\subsection{Causality, hyperbolic Fresnel polynomials and the Fresnel operator}\label{sub:causality}

It is well-known that the causal structure underlying Maxwell electrodynamics is given by a Lorentzian metric.
This may be explained by the fact that one can always choose a gauge, the Lorenz gauge, such that the field equations become manifestly hyperbolic with a principal symbol given by a Lorentzian metric.
This principal symbol can then be inverted everywhere except on its roots, \ie, the lightlike covectors, which is the basis for the well-posedness of the initial value problem for the field equations and the causal behaviour of the solutions of the theory (finite speed of propagation of disturbances).

In pre-metric electrodynamics there is typically no Lorentzian metric governing the causal behaviour of solutions.
A priori it is not even clear if the theory has a well-posed initial value problem and exhibits a causal behaviour.
Whether the theory is well-behaved in this sense is determined by the Fresnel polynomial, which is induced by the constitutive density $\chi$.
The importance of the Fresnel polynomial is that it plays a similar role in the field equations of pre-metric electrodynamics as the metric in Maxwell electrodynamics.
Namely, the points where the principal symbol of $P$~\eqref{eq:potential} cannot be inverted (in the sense of the previous section) are given by the roots of the Fresnel polynomial.

The foundation for the causal structures described in this section are standard results from the theory of linear partial differential equations with constant coefficients, as for example investigated in~\cite{Hoermander2}.
A modern general mathematical discussion on the relation between hyperbolic partial differential equations and causal structure can be found in~\cite{Khavkine:2014kya}.
It guarantees the existence of a causal structure which is a generalization of the usual Lorentzian causal structure if and only if the Fresnel polynomial is a hyperbolic polynomial.
Since in this article we aim for solution of the field equation with constant coefficients on the manifold $M = \RR^4$ as a first step towards the solution of the general case, we restrict our attention to constant Fresnel tensors densities.
\emph{That is, we will assume that the constitutive density~$\chi$ is given in a global Cartesian coordinate system where its components are constant, so that $\mathcal{G}$ as given by~\eqref{eq:fresnel_poly} is a hyperbolic polynomial independent of the global Cartesian coordinates chosen.}

The Fresnel polynomial $\mathcal{G}(k)$ is a fourth order homogeneous polynomial.
Abusing~\cite[Thm.~12.4.3]{Hoermander2} as a definition, we say that it is \emph{hyperbolic} at $x$ with respect to a covector $n$ if the map
\begin{equation*}
  \tau \mapsto \mathcal{G}(x, k + \tau n)
\end{equation*}
has only real roots for all real covectors~$k$.
Since we assumed $\mathcal{G}(x,k)=\mathcal{G}(k)$ there is no need to distinguish between hyperbolicity at a point $x$ and the global hyperbolicity of $\mathcal{G}$.
Each hyperbolicity covector $n$ belongs to an open convex cone\footnote{A cone in a vector space $V$ is a set $\Gamma \subset V$ such that $v \in V$ implies $\lambda v \in V$ for all $\lambda > 0$.}, the \emph{hyperbolicity cone} $\Gamma = \Gamma(n)$ of covectors with respect of which $\mathcal{G}(k)$ is also hyperbolic~\cite[Cor.~12.4.5]{Hoermander2}.
Such a cone should be understood to consist of `timelike' covectors and thereby defines a `time-orientation' for covectors.
The fact that we are working with a Fresnel polynomial which is independent of $x$ implements that $\Gamma$ is a hyperbolicity cone in each cotangent space of spacetime.
This means we can identify $\Gamma$ at all points of $M=\RR^4$.
Furthermore, we observe that hyperbolicity cones come in pairs:
If $n$ is a hyperbolicity covector, then so is $-n$~\cite[Thm.~12.4.1]{Hoermander2}; we set $-\Gamma = \Gamma(-n)$ for the corresponding opposite hyperbolicity cone.
A pair of hyperbolicity cones, a hyperbolicity double cone, plays the role of the future and past directed lightcones of the Lorentzian metric in Maxwell electrodynamics.

As stated above, our reason for studying hyperbolic Fresnel polynomials is the importance of the hyperbolicity property when solving differential equations that occur in relativistic physics.
Therefore we will study simultaneously to the causal notions defined by $\mathcal{G}(k)$ the fourth order partial differential operator $\mathcal{G}(\partial)$ defined as
\begin{equation}\label{eq:Gop}
  \mathcal{G}(\partial) \defn \mathcal{G}^{abcd} \partial_a \partial_b \partial_c \partial_d.
\end{equation}
Note that $\mathcal{G}(\partial)$ plays a crucial role when we construct the inverse of the field equations in Sect.~\ref{sub:inv}.
Moreover, $\mathcal{G}(\partial)$ defines an interesting partial differential field equation in itself.
It can be seen as a generalization of the wave operator and thus, if a mass term is added, leads to a generalization of the Klein--Gordon equation which is compatible with the dispersion relation dictated by the Fresnel polynomial.

It follows from~\cite[Thm.~12.5.1]{Hoermander2} that \emph{we can find to each hyperbolicity cone $\Gamma$ an inverse~$\mathcal{G}^{-1}_\Gamma$ of~$\mathcal{G}(\partial)$}, given for compactly supported $1$-densities $f$ by the operator
\begin{equation}\label{eq:Ginv}
  \mathcal{G}^{-1}_\Gamma f(x)
  \defn (2 \uppi)^{-4}\! \int_{\RR^4} \e^{\im (k - \im n) \cdot x}\, \frac{ \what{f}(k - \im n)}{\mathcal{G}(k - \im n)}\, \dif^4k
\end{equation}
by choosing any $n \in \Gamma$.
That is, the integral kernel of the inverse is given by
\begin{equation*}
  \mathcal{G}^{-1}_\Gamma(x, y) = \lim_{\varepsilon \searrow 0}\, (2 \uppi)^{-4}\! \int_{\RR^4} \frac{\e^{\im k \cdot (x - y)}}{\mathcal{G}(k - \im \varepsilon n)}\, \dif^4k,
\end{equation*}
where the limit is understood in the distributional sense.
The idea behind~\eqref{eq:Ginv} is that the hyperbolicity property allows a shifting of the integration contour into the complex, where no singularities of $\mathcal{G}^{-1}$ can be encountered.
Given a compactly supported $1$-density $f(x)$, a solution of
\begin{equation}\label{eq:Gdiffeq}
  \mathcal{G}(\partial) \varphi(x) = f(x)
\end{equation}
is given by $\varphi(x)=\mathcal{G}^{-1}_\Gamma f(x)$.
Due to the scalar density nature of $\mathcal{G}$, its inverse $\mathcal{G}^{-1}$ is also a scalar density but of weight with opposite sign.

The hyperbolicity cones of a hyperbolic polynomial give rise to an important cone structure for vectors which can be used to describe the support of solutions of the associated differential operator.
The dual cone $\Gamma^\circ$ is the closed convex cone of vectors $X$ such that $X(n) \geq 0$ for all $n \in \Gamma$:
\begin{equation*}
  \Gamma^\circ \defn \big\{ X \in T_xM \;\big|\; X(n) \geq 0 \textrm{ for all } n \in \Gamma \big\}.
\end{equation*}

In analogy to the causal sets $J_\pm(x)$ in Lorentzian geometry we define:
\begin{definition}
  The \emph{causal future of $x \in \RR^4$ with respect to $\Gamma$}, denoted by $J_\Gamma(x) \subset \RR^4$, is the closed convex cone with vertex at $x$ which consists of points that can be reached from $x$ by curves whose tangents lie in $\Gamma^\circ$.
  We also call the causal future of $x$ with respect to $-\Gamma$, denoted by $J_{-\Gamma}(x)$, the \emph{causal past of $x$ with respect to $\Gamma$}.
  The causal future (past) with respect to $\Gamma$ of a region $U \in \RR^4$  is defined as the union of the causal future (past) with respect to $\Gamma$ over all points of $U$:
  \begin{equation*}
    J_{\pm\Gamma}(U) \defn \bigcup_{x \in U} J_{\pm\Gamma}(x).
  \end{equation*}
\end{definition}

Applying this definition to the Fresnel polynomial and the associated differential operator~\eqref{eq:Gop}, we can state that the inverse $G^{-1}_\Gamma$~\eqref{eq:Ginv} has the support property~\cite[Thm.~12.5.1]{Hoermander2}
\begin{equation}\label{eq:Gsupp}
  \supp (\mathcal{G}^{-1}_\Gamma f) \subset J_\Gamma(\supp f).
\end{equation}
In other words, the maximum speed of propagation manifests itself in the set~$J_{\Gamma}(\supp f)$.

Turning the last definition on its head, we define:
\begin{definition}
  A set $U$ is called \emph{future compact with respect to $\Gamma$} (or \emph{$\Gamma$-future compact}) if
  \begin{equation*}
    U \cap J_\Gamma(x)
  \end{equation*}
  is compact for all $x \in \RR^4$.
  Similarly, $U$ is called \emph{past compact with respect to $\Gamma$} (or \emph{$\Gamma$-past compact}) if it is $-\Gamma$-future compact.
  If $U$ is both $\Gamma$-future and -past compact, we say that it is $\Gamma$-timelike compact.
\end{definition}
The notion of future and past compactness with respect to $\Gamma$ can be assigned to functions via their support.
Function spaces whose elements satisfy such support properties are denoted with a subscript $\Gamma\mathrm{fc}$ (for $\Gamma$-future compact), $\Gamma\mathrm{pc}$ (for $\Gamma$-past compact) or $\Gamma\mathrm{tc}$ (for $\Gamma$-timelike compact), \eg, we write $C^\infty_{\Gamma\mathrm{pc}}(\RR^4)$ for the space of $\Gamma$-past compact functions.

Not only are the solutions $\mathcal{G}^{-1}_\Gamma f$ of~\eqref{eq:Gdiffeq} supported in $J_\Gamma(\supp f)$, but $\mathcal{G}^{-1}_\Gamma f$ is in fact the only solution that is $\Gamma$-past compact.
Namely, it follows from~\cite[Thm.~8.6.9]{Hoermander1} that \emph{$\mathcal{G}^{-1}_\Gamma$ is the unique inverse of $\mathcal{G}(\partial)$ whose range is contained in the $\Gamma$-past compact functions}.

Let us explain how \emph{the domain of $\mathcal{G}^{-1}_\Gamma$ can be extended to $\Gamma$-past compact densities} `by duality'.
It follows from the assumption of constant coefficients, that \emph{$\mathcal{G}(\partial)$ is ``formally self-adjoint''}%
\footnote{%
  As in~\eqref{eq:P_selfadj}, $\mathcal{G}(\partial)$ is not formally self-adjoint in the usual sense since it is not a scalar operator but a scalar density.
  The difference to the usual self-adjointness of scalar partial differential operators is that we do not need an extra density factor in the integrals displayed.
}
in the sense
\begin{equation*}
  \int_{\RR^4} \big( \mathcal{G}(\partial) \varphi \big)\, \psi\, \dif^4x
  = \int_{\RR^4} \varphi\, \big( \mathcal{G}(\partial) \psi \big)\, \dif^4x
\end{equation*}
for all functions $\varphi, \psi$ such that $\supp \varphi \cap \supp \psi$ is compact.
As a consequence we find for all compactly supported densities $f, g$
\begin{align*}
  \int_{\RR^4} (\mathcal{G}^{-1}_\Gamma f)\, g\, \dif^4x
  &= \int_{\RR^4} (\mathcal{G}^{-1}_\Gamma f)\, (\mathcal{G}(\partial) \mathcal{G}^{-1}_{-\Gamma} g)\, \dif^4x
  = \int_{\RR^4} (\mathcal{G}(\partial) \mathcal{G}^{-1}_\Gamma f)\, (\mathcal{G}^{-1}_{-\Gamma} g)\, \dif^4x \\
  &= \int_{\RR^4} f\, (\mathcal{G}^{-1}_{-\Gamma} g)\, \dif^4x.
\end{align*}
Using this ``adjoint relation'', we continuously extend the domain of the inverse $\mathcal{G}^{-1}_\Gamma$ to $\Gamma$-past compact densities $f$ by setting
\begin{equation}\label{eq:ginvext}
  \int_{\RR^4} (\mathcal{G}^{-1}_\Gamma f)\, g\, \dif^4x = \int_{\RR^4} f\, (\mathcal{G}^{-1}_{-\Gamma} g)\, \dif^4x,
\end{equation}
for all compactly supported $g$, which defines $\mathcal{G}^{-1}_\Gamma f$ uniquely as a function in $C^\infty_{\Gamma\mathrm{pc}}(\RR^4)$.
Analogously we can extend the domain of $\mathcal{G}^{-1}_{-\Gamma}$ to $\Gamma$-future compact densities.
Note that $\mathcal{G}^{-1}_{\Gamma}$ cannot only act on scalar functions but also on $1$-forms $A$ or general tensorial objects, where its action then has to be understood componentwise $\mathcal{G}^{-1}_{\Gamma} A = \mathcal{G}^{-1}_{\Gamma} A_a(x)\, \dx^a$.

Later in Sect.~\ref{sub:sympl} we will briefly discuss the initial value problem for the field equation~\eqref{eq:potential} and thus need the concept of Cauchy surfaces.
The notions of causal past and causal future immediately yield such a definition:
\begin{definition}\label{def:cauchy}
  A hypersurface $\Sigma \subset \RR^4$ is called a \emph{Cauchy surface with respect to $\Gamma$} (or \emph{$\Gamma$-Cauchy surface}) if there exists a $1$-form $n$ which induces the distribution\footnote{Here we mean by `distribution' a subbundle $T\Sigma$ of the tangent bundle $TM$. The distribution induced by $n$ is given by $\ker n(x) = T_x\Sigma \subset T_xM$.} $T\Sigma \subset TM$ and $n(x) \in \Gamma$ for every $x \in \Sigma$.
  Moreover,
  \begin{equation*}
    J_{\Gamma}(\Sigma) \cup J_{-\Gamma}(\Sigma) = \RR^4,
  \end{equation*}
  \viz, every point of $\RR^4$ can be reached from $\Sigma$ by curves with tangents in $\pm\Gamma^\circ$.
\end{definition}

Often we will be concerned with solutions to equations whose restriction to a Cauchy surface as defined above is compactly supported.
Therefore we define a notion of spacelike compactness:
\begin{definition}
  A set $U$ is called \emph{spacelike compact with respect to $\Gamma$} (or \emph{$\Gamma$-spacelike compact}) if $U$ is closed and there exists a compact $K \subset \RR^4$ such that
  \begin{equation*}
    U \subset \big( J_\Gamma(K) \cup J_{-\Gamma}(K) \big).
  \end{equation*}
  In other words, for every $\Gamma$-Cauchy surface $\Sigma$ the intersection $U \cap \Sigma$ is compact.
\end{definition}
We say that a function $f$ is $\Gamma$-spacelike compact if this is true for its support and label function spaces of $\Gamma$-spacelike compact elements by a subscript $\Gamma\mathrm{sc}$.
For example, it follows from~\eqref{eq:Gsupp} that $\mathcal{G}^{-1}_{\pm\Gamma} f \in C^\infty_{\Gamma\mathrm{sc}}(\RR^4)$, the space of smooth $\Gamma$-spacelike compact functions, for any compactly supported density $f$.

All causality notions we introduced here and also the inverses are labelled by a hyperbolicity cone $\Gamma$ because, in general, there exist hyperbolic Fresnel polynomials which have more than one hyperbolicity double cone thus giving rise to inequivalent notions of `time'.
We address the involved subtleties for the physical viability of the theory while we go on.
It has already very generally be discussed in~\cite{Raetzel:2010je} that theories which lead to causal structures with hyperbolic polynomials leading to different inequivalent notions of time are problematic in their physical interpretation.

We now return to the path towards quantization of pre-metric electrodynamics and discuss the gauge freedom of the theory in more detail.

\subsection{Gauge fixing operators}\label{sub:gauge}

A very interesting and important object is the dual vector $\kappa^a(x, k)$ to each $k \in T^*_x M \otimes \CC$ which defines a gauge fixing $\kappa^a \what{A}_a = 0$ in momentum space, as discussed at the end of Sect.~\ref{sub:quasiinv}.
For the purpose of deriving the quasi-inverse $\wwhat{E}_{a}{}^{bcd}(x, k)$, the vector $\kappa(x, k)$ can be chosen freely as long as it is dual to $k$, \ie, it satisfies $k_a \kappa^a = 1$.
To associate to $\kappa$ a well-defined operator~$\vartheta$ we employ a definition via the Fourier transform on compactly supported functions $f$.
Assuming the poles of~$\kappa$ are determined by a hyperbolic polynomial, as it will be in the cases of interest below, we can define for each hyperbolicity cone $\Gamma$ and compactly supported $1$-form $A$, we have
\begin{equation*}
  (\vartheta_\Gamma A)(x) = -\im (2 \uppi)^{-4}\! \int_{\RR^4} \e^{\im (k - \im n) \cdot x} \kappa^a(x, k - \im n) \what{A}_a(k - \im n)\, \dif^4k,
\end{equation*}
so that the gauge fixing $\kappa^a \what{A}_a = 0$ is equivalent to $\vartheta_\Gamma A = 0$.
Moreover the same calculation shows that $\vartheta_{-\Gamma} A = 0$, simply change $n$ to $-n$ above.

Instead of taking up the difficult task of classifying all possible gauge choices, we will focus on two important cases and restrict to $\kappa$ which are position-independent.
The first case that we will look at is
\begin{subequations}\begin{align}
  \kappa^a(k) &= \frac{g^{ab} k_b}{g^{cd} k_c k_d} \label{eq:kappa_metric}\\
\intertext{and the second case is}
  \kappa^a(k) &= \frac{\mathcal{G}^{abcd} k_b k_c k_d}{\mathcal{G}(k)} \label{eq:kappa_fresnel},
\end{align}\end{subequations}
whenever the denominators are non-zero and with both $g^{ab}$ and $\mathcal{G}^{abcd}$ assumed constant, as we also did in the previous section.
We will further restrict the admissible $g^{ab}$ and $\mathcal{G}^{abcd}$ in the following paragraphs where we will discuss the two cases separately.
Observe that the first case is not the canonical gauge choice from viewpoint of pre-metric electrodynamics, since a canonical choice of the metric $g^{ab}$ is not available for every constitutive density $\chi$.
The second choice is always applicable and thus may be considered the canonical gauge choice of pre-metric electrodynamics.

\paragraph{Case 1.}
Suppose that $g$ is a Lorentzian metric with a timelike vector $n$ that defines a time-orientation.
It is well-known, see \eg~\cite{bar:2007}, that the d'Alembert operator $\Box = -g(\partial, \partial)$ associated to the Lorentzian metric $g$ possess unique retarded $\Box^{-1}_+$ and advanced $\Box^{-1}_-$ Green's operators.
For better agreement with the notation in the last section, we write $\Box^{-1}_\Gamma = \Box^{-1}_+$.
Then we can define $\vartheta_\Gamma : \Omega^1_{\Gamma\mathrm{pc}}(M) \to C^\infty_{\Gamma\mathrm{pc}}(M)$ as%
\footnote{\label{fn:1}To avoid confusion we mention that $g(\partial, A)=g^{ab}\partial_aA_b$ and $\mathcal{G}(\partial, \partial, \partial, A)=\mathcal{G}^{abcd}\partial_a\partial_b\partial_cA_d$ where $A$ cannot be interchanged with the $\partial$ in the arguments of $g$ and $\mathcal{G}$.}
\begin{align}\label{eq:theta_metric}
  \vartheta_\Gamma A &\defn -\Box^{-1}_\Gamma\big(g(\partial, A)\big).
\end{align}
Up to some technicalities, the symbol of $\vartheta_\Gamma$ in~\eqref{eq:theta_metric} is given by~\eqref{eq:kappa_metric}.

\paragraph{Case 2.}
Suppose that the Fresnel polynomial $\mathcal{G}(x, k)$ satisfies the assumptions of the previous Sect.~\ref{sub:causality}: it has constant coefficients and is hyperbolic.
Given a hyperbolicity cone $\Gamma$ with arbitrary $n \in \Gamma$, we can define $\vartheta_\Gamma : \Omega^1_{\Gamma\mathrm{pc}}(M) \to C^\infty_{\Gamma\mathrm{pc}}(M)$ as\footref{fn:1}
\begin{equation}\label{eq:theta_fresnel}
  \vartheta_\Gamma A \defn \mathcal{G}^{-1}_\Gamma\big(\mathcal{G}(\partial, \partial, \partial, A)\big).
\end{equation}
The symbol of $\vartheta_\Gamma$ in~\eqref{eq:theta_fresnel} is essentially given by~\eqref{eq:kappa_fresnel}.

\bigskip

In either case it is clear that \emph{$\vartheta_\Gamma$ inherits from $\kappa$ the duality property}
\begin{equation}\label{eq:thetadual}
  \vartheta_\Gamma \circ \dif = \id.
\end{equation}
Consequently the operators
\begin{equation}\label{eq:projop}
  \pi^{}_\Gamma{} \defn \mathord{\id} - \dif \circ \vartheta_\Gamma
  \quad\text{and}\quad
  \dif \circ \vartheta_\Gamma
\end{equation}
are projectors from $\Omega^1_{\Gamma\mathrm{pc}}(M)$ into itself; they play an essential role in the construction of the fundamental solution.
The kernel of $\dif \circ \vartheta_\Gamma$ consists of those $1$-forms $A$ that satisfy $\vartheta_\Gamma A = 0$.
We note that $\vartheta_\Gamma A = 0$ is a gauge condition specified by the choice of $\kappa$.
Namely, suppose that~$A'$ does not satisfy this condition, then
\begin{equation*}
  A = \pi_\Gamma A'
  = A' - \dif (\vartheta_\Gamma A')
  = A' + \dif \lambda
\end{equation*}
satisfies the gauge condition and differs from $A'$ by a gauge transformation.
Thus we see that \emph{the projector $\pi_\Gamma$ maps into the gauge-fixed $1$-forms of $\Gamma$-past compact support}.
Observe that the condition $\vartheta_\Gamma A = 0$ fixes the gauge completely (within the set of $\Gamma$-past compact $1$-forms) since a gauge transformation $A \mapsto A' = A + \dif \lambda$ with $\lambda \in C^\infty_{\Gamma\mathrm{pc}}(M)$ yields
\begin{equation}\label{eq:gaugefixc}
  \vartheta_\Gamma A'
  = \vartheta_\Gamma (A + \dif \lambda)
  = \lambda \neq 0.
\end{equation}

Instead of the gauge fixing on $\Gamma$-past compact $1$-forms $\vartheta_\Gamma A = 0$, it is possible to use alternatively, depending on the choice of $\vartheta_\Gamma$, the gauge conditions
\begin{equation*}
  g(\partial, A) = 0
  \quad\text{or}\quad
  \mathcal{G}(\partial, \partial, \partial, A) = 0.
\end{equation*}
These have the advantage that they can be applied independently of the support of the field~$A$.
However, for general support of $A$, they do not fix the gauge completely but leave the freedom of a gauge transformation $A \mapsto A' = A + \dif \lambda$ such that
\begin{equation*}
  \Box \lambda = 0
  \quad\text{or}\quad
  \mathcal{G}(\partial) \lambda = 0.
\end{equation*}
Solutions to these equations exist; in the second case it can be constructed from the solution of the inhomogeneous equation $\mathcal{G}(\partial) \varphi = f$ which we studied in~\eqref{eq:Ginv}.
They are never $\Gamma$-past or $\Gamma$-future compactly supported but may be $\Gamma$-spacelike compact.\footnote{Take the operator $\mathcal{G}^{-1}_\Gamma - \mathcal{G}^{-1}_{-\Gamma}$ to construct homogeneous solutions. See also the related construction of the Pauli--Jordan propagator in Sect.~\ref{sub:causp}.}

While the first gauge condition is the well-known \emph{Lorenz gauge} (sometimes also called Landau or Lorentz gauge), the second gauge condition is, to the knowledge of the authors, unknown in the literature; we shall call it the \emph{generalized Lorenz gauge}.
We chose this name because in the Lorentzian case, where the Fresnel tensor density is given by~\eqref{eq:maxwell_fresnel}, we find that the conditions
\begin{equation*}
  \mathcal{G}(\partial, \partial, \partial, A) = \Box\big(g(\partial, A)\big) = 0
  \quad\Leftrightarrow\quad
  g(\partial, A) = 0
\end{equation*}
are equivalent for $\Gamma$-past compact $1$-forms because there are no $\Gamma$-past compact solutions to the homogeneous equation $\Box \varphi = 0$.
For general $1$-forms the equivalence is not true.

Having clarified the gauge properties of the theory we are now able to write down the inverse of the field equations of pre-metric electrodynamics.

\subsection{Inverses of the field equation}\label{sub:inv}

We will now derive inverses, often called Green's operators or propagators, to the operator $P$ from~\eqref{eq:potential} under the assumption that the Fresnel polynomial $\mathcal{G}(k)$ is a constant coefficient hyperbolic polynomial, see Sect.~\ref{sub:causality}.
This is a first step towards the more difficult analysis of the general case of variable coefficients, which would be based on the analysis of the constant coefficient case by a perturbation argument.

As derived in Sect.~\ref{sub:quasiinv}, the Fresnel polynomial is central in the analysis of the principal symbol of the field equation of pre-metric electrodynamics.
Below we will see that our restriction to hyperbolic Fresnel polynomials leads a to theory of pre-metric electrodynamics that has a well-posed initial value problem and exhibits a causal behaviour.

Let $\kappa(k)$ be the canonical dual of pre-metric electrodynamics, given by~\eqref{eq:kappa_fresnel}, as described in the previous section.
In this section we will see that the map $E^\Gamma$ given by
\begin{equation}\label{eq:inv_symbl}
  (E^\Gamma J)_a(x) = (2 \uppi)^{-4}\! \int_{\RR^4} \e^{\im (k - \im n) \cdot x} \wwhat{E}_{a}{}^{bcd}(k - \im n) \what{J}_{bcd}(k - \im n)\, \dif^4k
\end{equation}
for all compactly supported $3$-forms $J$, is an inverse of $P$ with the support property
\begin{equation*}
  \supp(E^\Gamma J) \subset J_\Gamma(\supp J)
\end{equation*}
for a given hyperbolicity cone $\Gamma$ of $\mathcal{G}(k)$.
The ingredients to this inverse are a hyperbolicity covector $n \in \Gamma$, the quasi-inverse $\wwhat{E}$ obtained in~\eqref{eq:qinv3form} of Sect.~\ref{sub:quasiinv} and a suitable set of $3$-forms $J$ on which the map acts.

Decomposing the quasi-inverse $\wwhat{E}$ into its constituents~\eqref{eq:qinv3form}, we can define $E^\Gamma$ in terms of the operators constructed in the previous two sections:
In Sect.~\ref{sub:causality} we already constructed the operator corresponding to $\mathcal{G}(k)^{-1}$: it is the inverse $\mathcal{G}^{-1}_\Gamma$ of $\mathcal{G}(\partial)$ for some hyperbolicity cone~$\Gamma$ of $\mathcal{G}(k)$.
Then in Sect.~\ref{sub:gauge} we constructed operators $\vartheta_\Gamma$ corresponding to $\kappa$ in the projector $\pi_\Gamma = \id - \dif \circ \vartheta_\Gamma$~\eqref{eq:projop}; here we will only consider the canonical choice given by~\eqref{eq:theta_fresnel}.
The missing ingredient is the second order partial differential operator $Q(\partial) : \Omega^3(M) \to \Omega^1(M)$ given by
\begin{equation*}
  Q(\partial)_a{}^{bcd} \defn \frac{1}{3!} \frac{1}{8} \varepsilon^{ebcd} \varepsilon_{e c_1 a_1 a_2} \varepsilon_{a d_2 b_1 b_2} \chi^{a_1 c_1 b_1 d_1} \chi^{a_2 c_2 b_2 d_2} \partial_{d_1} \partial_{c_2}
\end{equation*}
corresponding to $Q_a{}^{bcd}(k)$ as defined in~\eqref{eq:defQ3form}.
Composing these operators we define
\begin{equation}\label{eq:inv}
  E^\Gamma \defn \pi_\Gamma \circ Q(\partial) \circ \mathcal{G}^{-1}_\Gamma
\end{equation}
acting on $\Gamma$-past compact, closed $3$-forms $J$ as $E^\Gamma J$ by letting $\mathcal{G}^{-1}_\Gamma$ act componentwise.

That the operator $E^\Gamma$ is well-defined follows from the properties of its constituents:
$\mathcal{G}^{-1}_\Gamma$ maps $\Omega^3_{\Gamma\mathrm{pc},\dif}(M)$ into $\Omega^3_{\Gamma\mathrm{pc}}(M)$, by its construction and its extension via the canonical pairing~\eqref{eq:ginvext}.
The operator $Q(\partial)$ maps $\Omega^3_{\Gamma\mathrm{pc}}(M)$ into $\Omega^1_{\Gamma\mathrm{pc}}(M)$.
Finally, $\pi_\Gamma$ maps non-gauge-fixed $1$-forms in $\Omega^1_{\Gamma\mathrm{pc}}(M)$ into gauge-fixed $1$-forms $\Omega^1_{\Gamma\mathrm{pc},\vartheta}(M)$.%
\footnote{Remember that the projector contains $\vartheta_\Gamma$ which is a composition of $\mathcal{G}^{-1}_\Gamma$ with the differential operator $\mathcal{G}(\partial, \partial, \partial, \cdot\,)$ (see~\eqref{eq:theta_fresnel}) so that it inherits domain and support properties from $\mathcal{G}^{-1}_\Gamma$.}
The subscript `$\vartheta$' on the $1$-form spaces indicates that the elements $A \in \Omega^1_{\Gamma\mathrm{pc},\vartheta}(M)$ satisfy the gauge condition $\vartheta_\Gamma A = 0$ and the subscript `$\dif$' on the $3$-form spaces their closedness.
These mappings can be visualized in the following diagram
\begin{equation*}
  \Omega^3_{\Gamma\mathrm{pc},\dif}(M)
  \overset{\mathcal{G}^{-1}_\Gamma}{\longrightarrow}
  \Omega^3_{\Gamma\mathrm{pc}}(M)
  \overset{Q(\partial)}{\longrightarrow}
  \Omega^1_{\Gamma\mathrm{pc}}(M)
  \overset{\pi_\Gamma}{\longrightarrow}
  \Omega^1_{\Gamma\mathrm{pc},\vartheta}(M)
\end{equation*}
All together we thus see that
\begin{equation}\label{eq:invmap}
  E^\Gamma : \Omega^3_{\Gamma\mathrm{pc},\dif}(M) \to \Omega^1_{\Gamma\mathrm{pc},\vartheta}(M).
\end{equation}
Note that $E^\Gamma$ contains only one projector $\pi_\Gamma$ while $\wwhat{E}$ defined in~\eqref{eq:qinv3form} contains two projectors.
The reason of this discrepancy is that we construct $E^\Gamma$ directly on closed $3$-forms so that the second projector is equivalent to the identity.
If we consider this, one can see that~\eqref{eq:inv_symbl} gives~\eqref{eq:inv} and~\eqref{eq:invmap} (after a proper extension of the operator).

\emph{The most important property of $E^\Gamma$ is that it is an inverse of $P$ acting on gauge-fixed $\Gamma$-past compact $1$-forms.}
The calculations done in~\eqref{eq:fourier_inv1} and~\eqref{eq:fourier_inv2} carry over directly to the corresponding operators:
\begin{subequations}\begin{align}
  P(E^\Gamma J)
  &= \big(P \circ \pi_\Gamma \circ Q(\partial) \circ \mathcal{G}^{-1}_\Gamma \big) J
  = \big(P \circ Q(\partial) \circ \mathcal{G}^{-1}_\Gamma \big) J
  = J, \label{eq:invj} \\
\intertext{when acting on closed $3$-forms $J$, and}
  E^\Gamma(P A)
  &= \big(\pi_\Gamma \circ Q(\partial) \circ \mathcal{G}^{-1}_\Gamma \circ P \big) A
  = \pi_\Gamma A=A, \label{eq:inva}
\end{align}\end{subequations}
when acting on gauge-fixed $1$-forms $A$.
Since $\pi_\Gamma$ is a projector into the gauge-fixed $1$-forms, it follows that $E^\Gamma$, considered as the map~\eqref{eq:invmap}, is an inverse of $P$.
A direct consequences of the inverse property of $E^\Gamma$ and~\eqref{eq:P_selfadj} is
\begin{equation}\label{eq:adj}
  \int_M E^\Gamma J \wedge K
  = \int_M E^\Gamma J \wedge P E^{-\Gamma} K
  = -\int_M P E^\Gamma J \wedge E^{-\Gamma} K
  = -\int_M J \wedge E^{-\Gamma} K,
\end{equation}
which demonstrates the ``adjoint relation'' between $E^\Gamma$ and $E^{-\Gamma}$ on compactly supported, closed $3$-forms $J, K$ with respect to their canonical pairing.
Concerning the gauge freedom of the theory, we also see immediately from~\eqref{eq:inva} that $E^\Gamma$ is not an inverse on non-gauge-fixed $1$-forms but only an inverse up to a gauge transformation
\begin{equation}\label{eq:invnongauge}
  E^\Gamma(P A) = A - \dif (\vartheta_\Gamma A) = A + \dif \lambda.
\end{equation}

To demonstrate in more detail that \emph{the range of $E^\Gamma$ are the gauge-fixed $1$-forms}, we apply the gauge fixing operator $\vartheta_\Gamma$ and find
\begin{equation*}
  \vartheta_\Gamma(E^\Gamma J) = \big( \vartheta_\Gamma \circ \pi_\Gamma \circ Q(\partial) \circ \mathcal{G}^{-1}_\Gamma \big) J = 0
\end{equation*}
because $\vartheta_\Gamma \circ \pi_\Gamma = 0$ by~\eqref{eq:thetadual}.
In virtue of~\eqref{eq:gaugefixc}, $A = E^\Gamma J$ is completely gauge-fixed since a gauge transformed $A' = A + \dif \lambda$, with $\lambda \in C^\infty_{\Gamma\mathrm{pc}}(M)$, would no longer solve the gauge condition.
Yet this is not the only condition that the solutions satisfy.
Observe that, due to $\mathcal{G}(\partial, \partial, \partial, \pi_\Gamma\,\cdot\,) = 0$ for our choice of $\vartheta_\Gamma = \mathcal{G}^{-1}_\Gamma \mathcal{G}(\partial, \partial, \partial, \cdot\,)$, the generated solutions $E^\Gamma J$ also satisfy the genearlized Lorenz gauge
\begin{equation}\label{eq:weakgauge}
  \mathcal{G}(\partial, \partial, \partial, E^\Gamma J) = \mathcal{G}\big(\partial, \partial, \partial, (\pi_\Gamma \circ Q(\partial) \circ \mathcal{G}^{-1}_\Gamma) J\big) = 0.
\end{equation}

Summing up, the operator $E^\Gamma$ constructed above can be used to obtain completely gauge-fixed (co)vector potentials which solve the inhomogeneous field equations of pre-metric electrodynamics.

If we do not care about the precise range of $E^\Gamma$, \viz, the precise gauge condition satisfied by the (co)vector potential, we can even drop the gauge fixing projector and use instead of~$E^\Gamma$ in~\eqref{eq:inv}
\begin{equation}\label{eq:inv_feynman}
  D^\Gamma \defn Q(\partial) \circ \mathcal{G}^{-1}_\Gamma.
\end{equation}
The $1$-forms generated with this operator would be gauge equivalent to the ones obtained with $E^\Gamma$.
In the case of Maxwell electrodynamics, we find, using~\eqref{eq:maxwell_Q},
\begin{equation}\label{eq:maxwell_feynman}
  D^\Gamma{}_a{}^{bcd} = (g_{ae} + \partial_a \partial_e) \varepsilon^{ebcd} \Box^{-1}_\Gamma = g_{ae} \varepsilon^{ebcd} \Box^{-1}_\Gamma,
\end{equation}
where the second equality holds because of our restriction of the domain to closed $3$-forms.
The rightmost side of~\eqref{eq:maxwell_feynman} is known as the Green's operator in the so-called Feynman gauge.

Instead of choosing~\eqref{eq:theta_fresnel} as gauge fixing operator, we could have also chosen~\eqref{eq:theta_metric} in~\eqref{eq:inv} given that the metric $g$ has the $\Gamma$ as a hyperbolicity cone.
We can even relax this requirement and only demand that $g$ has a hyperbolicity cone $\Gamma'$ which is contained in $\Gamma$.
In the opposite situation where $\Gamma$ is contained in $\Gamma'$ it is still possible to construct an inverse if one restricts its domain to compactly supported $3$-forms.
For the same reason it is possible to choose for the construction of the gauge fixing operator via~\eqref{eq:theta_fresnel} a different Fresnel tensor density as long as it possesses a hyperbolicity cone that overlaps with $\Gamma$.
Nevertheless, all these choices are usually not very natural and we will abstain from discussing them any further.
However, in some situations such as for uniaxial crystals, which we already mentioned above and discuss in more detail in App.~\ref{app:Uniaxial}, the fourth order Fresnel polynomial is a product of two quadratic metric polynomials.
In these cases there is a canonical choice of a metric gauge condition available and can be used.

\subsection{Pauli--Jordan propagators}\label{sub:causp}

The inverses constructed in the previous section generate solutions to the inhomogeneous field equation~\eqref{eq:potential} with constant coefficients.
As we have seen, there exists one inverse $E^\Gamma$ for each hyperbolicity cone $\Gamma$ of the Fresnel polynomial.
The theory of hyperbolic polynomials guarantees that hyperbolicity cones come in pairs: if $\Gamma$ is a hyperbolicity cone, so is the opposite cone $-\Gamma$.
These hyperbolicity double cones give rise to the causal notions that we introduced in Sect.~\ref{sub:causality}.
Thus, whenever the Fresnel polynomial is hyperbolic, \ie, for all constitutive laws for which we constructed the fundamental solutions in the previous section, there exists the \emph{Pauli--Jordan propagator}
\begin{equation}\label{eq:pjprop}
  \upDelta^\Gamma \defn E^{-\Gamma} - E^\Gamma.
\end{equation}
We immediately see that \emph{ $\upDelta^\Gamma$ generates solutions to the homogeneous field equations $P A = 0$} because
\begin{equation*}
  P (\upDelta^\Gamma J) = P (E^{-\Gamma} J) - P (E^\Gamma J) = 0
\end{equation*}
as a consequence of equation~\eqref{eq:invj}.
By construction $\upDelta^\Gamma J$ has support in $J_\Gamma(\supp J) \cup J_{-\Gamma}(\supp J)$, \ie, in the union of the causal $\Gamma$-future and the causal $\Gamma$-past of the support of~$J$.
For this reason $\upDelta^\Gamma$ is sometimes also called the \emph{causal propagator}.

Observe that we cannot claim that solutions of the homogeneous field equation $\upDelta^\Gamma J$ satisfy a gauge condition like $\vartheta_\Gamma (\upDelta^\Gamma J) = 0$ since $\upDelta^\Gamma J$ cannot be $\Gamma$-past compact.
In any case, from~\eqref{eq:weakgauge} it is clear that $\upDelta^\Gamma J$ satisfies what we called the generalized Lorenz gauge in Sect.~\ref{sub:gauge}, namely,
\begin{equation*}
  \mathcal{G}(\partial, \partial, \partial, \upDelta^\Gamma J) = 0.
\end{equation*}

In Maxwell electrodynamics, once the gauge has been fixed, the causal propagator is unique (up to a sign) because there exists only one pair of hyperbolicity cones.
However, in pre-metric electrodynamics there exist Fresnel tensors which have several pairs of hyperbolicity cones $\Gamma_i$, $-\Gamma_i$.
In these cases we can associate one Pauli--Jordan propagator~$\upDelta^{\Gamma_i}$ to each such pair.
We would like to remark here, that it is not clear if the hyperbolic polynomials which posses several hyperbolicity double cones can be interpreted physically.
The non-uniqueness of the Pauli--Jordan propagator causes several problems for the classical and the quantum theory.
It gives rise to a natural \mbox{(pre-)symplectic} form, see Sect.~\ref{sub:sympl}, which is (of course) closely connected to a Hamiltonian formulation of the theory.
In case of multiple inequivalent propagators, one would generically expect that no (unique) Hamiltonian formulation exists; there would be one Hamiltonian formulation for each hyperbolicity double cone.
It is doubtful that these can be interpreted consistently.
However, there exists a vast variety of constitutive laws which lead to a hyperbolic Fresnel polynomial that possesses only one hyperbolicity double cone.
The dispersion relations which describe linear dielectric and permeable media satisfy this condition~\cite{Perlick}.
Among them are the dispersion relations of uniaxial crystals, as we demonstrate explicitly in App.~\ref{app:Uniaxial}.
In Sect.~\ref{sec:quant} we will only quantize theories with a single hyperbolicity double cone.
Note that constitutive laws yielding multiple hyperbolicity double cones cannot be bihyperbolic\footnote{\cite{Raetzel:2010je} call a principal symbol bihyperbolic if it is hyperbolic and possesses a certain dual symbol that is also hyperbolic.} as defined in~\cite{Raetzel:2010je}, where it is argued that only bihyperbolic theories can be considered physical.

The following properties of the Pauli--Jordan propagator are independent of the number of hyperbolicity double cones:

The domain of the Pauli--Jordan propagator is, by its construction from the Green's operators~$E^\Gamma$ and~$E^{-\Gamma}$, the intersection of their domains, \ie, the $\Gamma$-timelike compact, closed $3$-forms.
Albeit their compactness to the past and the future with respect to $\Gamma$, they may have non-compact $\Gamma$-spacelike support.
The range of the propagator is contained in the space of $1$-forms which satisfy the homogeneous field equation.
When restricted to $3$-forms $J$ with compact support, the resulting (co)vector potential will be $\Gamma$-spacelike compactly supported.
These support properties follow from the union of the support of $E^\Gamma J$ and $E^{-\Gamma} J$ discussed in the preceding section.

\emph{Every solution $A$ of the homogeneous field equation $P A = 0$ is gauge-equivalent to a solution $A' = \upDelta^\Gamma J$ for some $\Gamma$-timelike compact, closed $3$-form $J$}.
To see this, let $\Psi$ be a $\Gamma$-past compact function such that $(1 - \Psi)$ is $\Gamma$-future compact, \viz, there exist $\Gamma$-Cauchy surfaces~$\Sigma$ and~$\Sigma'$ such that $\Psi(J_\Gamma(\Sigma)) = 1$ and $\Psi(J_{-\Gamma}(\Sigma')) = 0$.
We can use $\Psi$ to decompose $A$ into the $\Gamma$-past compact $A^+ = \Psi A$ and $\Gamma$-future compact $A^- = (1 - \Psi) A$ so that $A = A^+ + A^-$.
Observe that $J = P A^+ = -P A^- = P(\Psi A)$ is only supported in a $\Gamma$-timelike compact set because it can only be supported where $\Psi$ is non-constant since we assumed $P A = 0$.
Using $J$, we find a solution $A' = \upDelta^\Gamma J$ to the homogeneous field equation.
From~\eqref{eq:inva} and~\eqref{eq:invnongauge} we see that $A'$ and $A$ are gauge-equivalent, because $E^\Gamma$ and $E^{-\Gamma}$ are inverses up to a gauge transformation~\eqref{eq:invnongauge}.
As a corollary to this statement it is evident that \emph{every spacelike compact solution of the homogeneous field equation $P A = 0$ is gauge-equivalent to a solution $A' = \upDelta^\Gamma J$ for some compactly supported,\footnote{For a spacelike compact solution $A$ the support of $J = P A^+$ is not only confined between the two Cauchy surfaces $\Sigma$ and $\Sigma'$, where $\Psi$ is not constant, but also $\Gamma$-spacelike compact and thus compact.} closed $3$-form $J$}.

Most properties above can nicely be summarised in the following two exact sequences, \cf~\cite[Thm.~3.4.7]{bar:2007}.
For $\Gamma$-timelike compact $A$ we have
\begin{equation*}
  0 \longrightarrow \Omega^1_{\Gamma\mathrm{tc},\vartheta}(M) \overset{P}{\longrightarrow} \Omega^3_{\Gamma\mathrm{tc},\dif}(M) \overset{\upDelta^\Gamma}{\longrightarrow} \Omega^1(M) \overset{P}{\longrightarrow} \Omega^3_\dif(M) \longrightarrow 0,
\end{equation*}
while for compact $A$ the exact sequence is
\begin{equation*}
  0 \longrightarrow \Omega^1_{c,\vartheta}(M) \overset{P}{\longrightarrow} \Omega^3_{c,\dif}(M) \overset{\upDelta^\Gamma}{\longrightarrow} \Omega^1_{\Gamma\mathrm{sc}}(M) \overset{P}{\longrightarrow} \Omega^3_{\Gamma\mathrm{sc},\dif}(M) \longrightarrow 0.
\end{equation*}
Observe that in the second step we used that $\upDelta^\Gamma(P A) = 0$ for $\Gamma$-timelike compact, gauge-fixed $A$ because $\vartheta_\Gamma A = 0 = \vartheta_{-\Gamma} A$.
For a non-gauge-fixed $\Gamma$-timelike compact $1$-form $A$, the Pauli--Jordan propagator generates a pure gauge solution
\begin{equation}\label{eq:PJnon-gauge}
  \upDelta^\Gamma(P A) = E^{-\Gamma}(P A) - E^\Gamma(P A) = \dif (\vartheta_{\Gamma} A - \vartheta_{-\Gamma} A) = \dif \lambda,
\end{equation}
see~\eqref{eq:invnongauge}.

In addition to generating the solution of the homogeneous field equations the Pauli--Jordan propagator enables us to construct a symplectic structure on the space of solutions.

\subsection{Symplectic structure and the classical phase space}\label{sub:sympl}

In the last section we used the inverses of Sect.~\ref{sub:inv} to construct a Pauli--Jordan propagator for each hyperbolicity double cone.
In this section we will use the propagator to classify the space of solutions of the homogeneous field equations of pre-metric electrodynamics corresponding to this hyperbolicity double cone and equip it with a natural symplectic structure.

Consider on $\Omega^3_{c,\dif}(M)$ the bilinear form
\begin{equation}\label{eq:sympl}
  \sigma^\Gamma(J, K) \defn \int_M J \wedge \upDelta^\Gamma K.
\end{equation}
It follows directly from the ``adjointness properties'' of $E^\Gamma$, see~\eqref{eq:adj}, that it is skew-symmetric
\begin{equation*}
  \sigma^\Gamma(J, K)
  = \int_M J \wedge \upDelta^\Gamma K
  = \int_M \upDelta^\Gamma J \wedge K
  = -\int_M K \wedge \upDelta^\Gamma J
  = -\sigma^\Gamma(K, J).
\end{equation*}
Therefore it is \emph{a pre-symplectic form on the space of compactly supported, closed $3$-forms}.
It degenerates on all $3$-forms that are given by $P A$ for $A \in \Omega^1_c(M)$, since
\begin{equation*}
  \sigma^\Gamma(J, P A)
  = \int_M J \wedge \upDelta^\Gamma (P A)
  = \int_M J \wedge \dif \lambda
  = \int_M \lambda\, \dif J
  = 0,
\end{equation*}
by~\eqref{eq:PJnon-gauge}, Stokes' theorem and the fact that $J$ is closed.
Thus $\sigma^\Gamma$ is degenerate and not a symplectic form which makes $(\Omega^3_{c,\dif}(M), \sigma^\Gamma)$ a pre-symplectic space -- it may be called the \emph{off-shell phase space} of the theory.
But at the same time this implies that $\sigma^\Gamma$ is defined independently of the gauge choice which enters implicitly via $\pi_{\pm\Gamma}$ in $\upDelta^\Gamma$.

It is now not difficult to show that the kernel of $\sigma^\Gamma$ is given by $P \Omega^1_c(M)$ so that (omitting the composition with the quotient map) \emph{$\sigma^\Gamma$ can be turned into a symplectic form on the quotient space}
\begin{equation*}
  \mathfrak{S}^*_{\Gamma\mathrm{sc}} \defn \Omega^3_{c,\dif}(M) \big/ P \Omega^1_c(M);
\end{equation*}
we call $(\mathfrak{S}^*_{\Gamma\mathrm{sc}}, \sigma^\Gamma)$ the \emph{on-shell phase space}.
This space can be identified with the space of solutions of the homogeneous field equations $\mathfrak{S}_{\Gamma\mathrm{sc}}$ induced by applying~$\upDelta^\Gamma$ to representatives of $\mathfrak{S}^*_{\Gamma\mathrm{sc}}$, \ie,
\begin{equation*}
  \mathfrak{S}_{\Gamma\mathrm{sc}} \defn \upDelta^\Gamma \mathfrak{S}^*_{\Gamma\mathrm{sc}} \subset \Omega^1_{\Gamma\mathrm{sc}}(M).
\end{equation*}
Clearly this is just a subspace of the whole space of solutions of the homogeneous field equations: $\mathfrak{S}_{\Gamma\mathrm{sc}}$ contains only one representative of each gauge equivalence class of $\Gamma$-spacelike compact solutions.

Also $\mathfrak{S}_{\Gamma\mathrm{sc}}$ can be equipped with a natural symplectic form.
Let $\Sigma$ be a arbitrary $\Gamma$-Cauchy surface and $A, B \in \mathfrak{S}_{\Gamma\mathrm{sc}}$, then we define
\begin{equation}\label{eq:sympl2}
  \varsigma^\Gamma(A, B)
  \defn \int_\Sigma (A \wedge \conlaw \dif B - B \wedge \conlaw \dif A).
\end{equation}
To see that this definition is independent of the $\Gamma$-Cauchy surface chosen, note that the exterior derivative applied to the integrand is zero, so that Stokes' theorem can be applied.
Using again Stokes' theorem, it can also be shown that $\varsigma^\Gamma$ is gauge-invariant when operating on \emph{any} solutions which are spacelike compact with respect to $\Gamma$.
Note that the symplectic form $\varsigma^\Gamma$ is equivalent to the ``charge'' of~\cite[Eq.~(42)]{Rivera:2011rx} in their choice of gauge and constitutive law.
We will now show that $(\mathfrak{S}^*_{\Gamma\mathrm{sc}}, \sigma^\Gamma)$ and $(\mathfrak{S}_{\Gamma\mathrm{sc}}, \varsigma^\Gamma)$ are indeed equivalent.

Given a $\Gamma$-Cauchy surface $\Sigma$, we can split the spacetime into $M = \Sigma^+ \cup \Sigma \cup \Sigma^-$, where $\Sigma^+$ is past compact and $\Sigma^-$ is future compact with respect to $\Gamma$.
Then, we can write
\begin{equation*}
  \sigma^\Gamma(J, K)
  = \int_M J \wedge \upDelta^\Gamma K
  = \int_M J \wedge B
  = \int_{\Sigma^+} J \wedge B + \int_{\Sigma^-} J \wedge B
\end{equation*}
For both integrals on the right-hand side we calculate
\begin{equation*}
  \int_{\Sigma^\pm} J \wedge B
  = \int_{\Sigma^\pm} \dif \conlaw \dif A^\mp \wedge B
  = \int_{\Sigma^\pm} \dif (B \wedge \conlaw \dif A^\mp - A^\mp \wedge \conlaw \dif B)
  = \pm \int_\Sigma (A^\mp \wedge \conlaw \dif B - B \wedge \conlaw \dif A^\mp),
\end{equation*}
where we set $\dif \conlaw \dif A^\pm = P(E^{\pm\Gamma} J) = J$, used Stokes' theorem, the symmetry of the constitutive law $\conlaw \dif B \wedge \dif A = \dif B \wedge \conlaw \dif A$ and the fact that $B \in \mathfrak{S}_{\Gamma\mathrm{sc}}$.
The sign in the last step occurs due to the relative induced orientation of the boundaries of $\partial \Sigma^+$ and $\Sigma = \partial \Sigma^-$.
Adding the results for $\Sigma^+$ and~$\Sigma^-$, we conclude that
\begin{equation*}
  \sigma^\Gamma(J, K)
  = \varsigma^\Gamma(\upDelta^\Gamma J, \upDelta^\Gamma K)
  = \varsigma^\Gamma(A, B).
\end{equation*}
\emph{Thus we can describe the phase space in terms of (equivalence classes of) currents $\mathfrak{S}^*_{\Gamma\mathrm{sc}}$ with symplectic form $\sigma$ or we can use the space of solutions $\mathfrak{S}_{\Gamma\mathrm{sc}}$ with symplectic form $\varsigma$}.

Actually, since~\eqref{eq:sympl2} only contains the Cauchy data for the solutions $A, B$, namely the pullback of $A, B$ and $\conlaw \dif A, \conlaw \dif B$ to the Cauchy surface, we can uniquely identify each solution in $\mathfrak{S}_{\Gamma\mathrm{sc}}$ with its Cauchy data.
This implies that we equivalently define the on-shell phase space in terms of the space of Cauchy data.
We remark that the pullback of $\conlaw \dif A$ and $\conlaw \dif B$ to the $\Gamma$-Cauchy surface are the canonical momenta of $A$ and $B$ at $\Sigma$.
We could have derived the same expression for the canonical momenta from the action of pre-metric electrodynamics
\begin{align}\label{eq:action}
  S[A] = \frac12 \int_M \dif A \wedge \conlaw \dif A = \frac12 \int_M \chi^{abcd} (\partial_a A_b) (\partial_c A_d)\, \dif^4 x
\end{align}
but we will not follow that approach here.

Finally, we point out that from the point of view of the Poisson geometry of the solution space one should call~\eqref{eq:sympl} Poisson bivector and~\eqref{eq:sympl2} symplectic form.
In this setting the Poisson bivector~\eqref{eq:sympl} acts on $\mathfrak{S}^*_{\Gamma\mathrm{sc}}$, which may be identified with the cotangent space of the solution space, and the symplectic form~\eqref{eq:sympl2} acts on $\mathfrak{S}_{\Gamma\mathrm{sc}}$, which may be identified with the tangent space of solution space.
Of course, since we consider a linear equation, $\mathfrak{S}_{\Gamma\mathrm{sc}}$ coincides with the solution space.
We refer to~\cite{Khavkine:2014kya} for an extensive discussion.

\subsection{The energy momentum of the electromagnetic field}\label{sub:energydensity}

In order to construct quantum states for the quantum field theory to be developed in the next section (Sect.~\ref{sec:quant}), we employ a positive inner product on the space of solutions of the homogeneous field equations.
A good candidate for such a function is the energy density of the electromagnetic field, which also leads to the desired inner product.

The axiomatic approach to electrodynamics by Hehl and Obukhov~\cite{Hehl} leads to the following covector-valued $3$-form, which is interpreted as \emph{kinematic energy-momentum} of the electromagnetic field
\begin{equation*}
  T_N\defn\frac{1}{2}\big(F\wedge(\iprod{N} H)-H\wedge(\iprod{N} F)\big),
\end{equation*}
where $N$ is a vector field.
$T_N$ is called the kinematic energy-momentum of the field since it is basically the potential which generates the Lorentz force acting on a particle travelling along an integral curve of $N$.

Since we also allow for complex solutions, it is necessary to `complexify' the energy-momentum.
Moreover, we can rewrite it in terms of the potential with help of the field equation and the constitutive law~\eqref{eq:conlawcoord}.
We denote the complexified energy-momentum $3$-form with the same symbol
\begin{equation*}
  T_N(A)\defn\frac{1}{2}\big(\dif\conj A\wedge(\iprod{N} \conlaw \dif A)-\conlaw \dif \conj A\wedge(\iprod{N} \dif A)\big).
\end{equation*}

The importance of this energy-momentum lies in the fact that it generates conservation laws and conserved quantities of the theory, when evaluated on the space of solutions of the homogeneous field equations.
For solutions of the homogeneous field equations~\eqref{eq:genlin3form} the exterior differential yields
\begin{equation*}
  2\,\dif T_N=\dif \conj A \wedge \mathcal{L}_N \conlaw \dif A -\conlaw \dif  \conj A\wedge \mathcal{L}_N\dif A=\dif \conj A \wedge \mathcal{L}_N(\conlaw)(\dif A).
\end{equation*}
Thus we find that for \emph{generalized Killing vector fields} $N$, \ie, vector fields that satisfy $\mathcal L_N \conlaw = 0$, $\dif T_N$ vanishes.
In the case when $N$ is the tangent vector field of an observer as defined in~\cite{Raetzel:2010je}, one can interpret $T_N$ as energy-momentum and $n \wedge T_N$, for $n$ being dual to $N$ (\ie, $N(n)=1$), as energy density associated to $A$ as measured by an observer flowing along $N$.

To analyse the positivity properties of the energy density, we express
\begin{equation*}
  \rho \defn \frac{1}{4!} \varepsilon^{abcd} (n \wedge T_N)_{abcd}
\end{equation*}
in terms of the field strength $F = \dif A$ in local coordinates
\begin{align*}
  n \wedge T_N
  &= \frac12 n \wedge \big( \conj{F} \wedge (\iprod{N} \conlaw F) - \conlaw \conj{F} \wedge (\iprod{N} F) \big) \\
  &= \frac12 n \wedge \big( \iprod{N} (\conj{F} \wedge \conlaw F) - (\iprod{N} \conj{F}) \wedge \conlaw F - \conlaw \conj{F} \wedge (\iprod{N} F) \big) \\
  &= \frac12 \big( \conj{F} \wedge \conlaw F - n \wedge (\iprod{N} \conj{F}) \wedge \conlaw F - \conlaw \conj{F} \wedge n \wedge (\iprod{N} F) \big) \\
  &= \frac18 \chi^{abcd} \big( \conj{F}_{ab} F_{cd} - 2 n_a N^e \conj{F}_{eb} F_{cd} - 2 \conj{F}_{ab} n_c N^e F_{ed} \big)\, \dx^0\! \wedge \dx^1\! \wedge \dx^2\! \wedge \dx^3
\end{align*}
and represent $\chi^{abcd}$ as a symmetric $6 \times 6$ matrix in the following way:
Let $\{e_a\}_{a=0}^3$ be a basis of the tangent spaces of spacetime with $e_0=N$ and, since $n$ is dual to $N$, $e_\alpha(n)=0$ with $\alpha=1,2,3$.
We can construct a basis $\{E_A\}_{A=1}^6$ on the six dimensional space of bi-vectors, that is the space dual to the $2$-form space on spacetime, by taking all possible pairwise wedge products
\begin{equation*}
  E_\alpha=N \wedge e_\alpha\ (\alpha=1,2,3),\quad E_4=e_2\wedge e_3,\quad E_5=e_3\wedge e_1,\quad E_6=e_1\wedge e_2.
\end{equation*}
In this basis $\chi^{abcd}$ is composed out of three matrices $X,Y,Z$, where $X$ and $Y$ are symmetric, and assumes the following form
\begin{equation*}
  (\chi^{AB})
  = \left(\begin{array}{c|c} \!\!(X^{\alpha\beta})\! & \!(Z^{\alpha\mathfrak{b}})\!\! \\ \hline \!\!(Z^{\mathfrak{a}\beta})\! & \!(Y^{\mathfrak{a}\mathfrak{b}})\!\! \end{array}\right)
  = \left(\begin{array}{ccc|ccc}
  \chi^{0101} & \chi^{0102} & \chi^{0103} & \chi^{0123} & \chi^{0131} & \chi^{0112} \\
  \chi^{0201} & \chi^{0202} & \chi^{0203} & \chi^{0223} & \chi^{0231} & \chi^{0212} \\
  \chi^{0301} & \chi^{0302} & \chi^{0303} & \chi^{0323} & \chi^{0331} & \chi^{0312} \\
  \hline
  \chi^{2301} & \chi^{2302} & \chi^{2303} & \chi^{2323} & \chi^{2331} & \chi^{2312} \\
  \chi^{3101} & \chi^{3102} & \chi^{3103} & \chi^{3123} & \chi^{3131} & \chi^{3112} \\
  \chi^{1201} & \chi^{1202} & \chi^{1203} & \chi^{1223} & \chi^{1231} & \chi^{1212}
  \end{array}\right),
\end{equation*}
where $\alpha,\beta=1,2,3$ and $\mathfrak{a},\mathfrak{b}=4,5,6$ label the different parts of the $E_A$ basis.
Therefore the energy density can be written as
\begin{align*}
  2 \rho
  &= \begin{pmatrix} F_\alpha \\ F_{\mathfrak{a}} \end{pmatrix}^{\!*}
  \begin{pmatrix} X^{\alpha\beta} & Z^{\alpha\mathfrak{b}} \\ Z^{\mathfrak{a}\beta} & Y^{\mathfrak{a}\mathfrak{b}} \end{pmatrix}
  \begin{pmatrix} F_\beta \\ F_{\mathfrak{b}} \end{pmatrix}
  - \begin{pmatrix} F_\alpha \\ 0 \end{pmatrix}^{\!*}
  \begin{pmatrix} X^{\alpha\beta} & Z^{\alpha\mathfrak{b}} \\ 0 & 0 \end{pmatrix}
  \begin{pmatrix} F_\beta \\ F_{\mathfrak{b}} \end{pmatrix}
  - \begin{pmatrix} F_\alpha \\ F_{\mathfrak{a}} \end{pmatrix}^{\!*}
  \begin{pmatrix} X^{\alpha\beta} & 0 \\ Z^{\mathfrak{a}\beta} & 0 \end{pmatrix}
  \begin{pmatrix} F_\beta \\ 0 \end{pmatrix} \\
  &= -X^{\alpha\beta}\conj{F}_{\alpha}F_{\beta} + Y^{\mathfrak{a}\mathfrak{b}}\conj{F}_{\mathfrak{a}}F_{\mathfrak{b}}.
\end{align*}
From this expression we find that \emph{the energy density an observer associates to the field in pre-metric electrodynamics is positive if and only if}
\begin{equation}\label{eq:positive_constlaw}
  -X^{\alpha\beta}\conj{F}_{\alpha}F_{\beta} + Y^{\mathfrak{a}\mathfrak{b}}\conj{F}_{\mathfrak{a}}F_{\mathfrak{b}} > 0
\end{equation}
for non-vanishing $F$, hence for constitutive laws $\chi$ for which the matrix $X$ is negative definite and the matrix $Y$ is positive definite.
Since this positivity property of $\rho$ will be essential for us in the construction of a state for the quantized theory, we will restrict to constitutive laws with this property.
A similar requirement for the quantization of pre-metric electrodynamics was derived in~\cite[App.]{Rivera:2011rx}.
There it is shown that the so-called bihyperbolic and energy-distinguishing area metrics, which correspond to our constitutive densities, have the property~\eqref{eq:positive_constlaw}.

While a sensible free classical theory should have a positive energy density as guaranteed by the conditions above, this positivity will also be crucial for the construction of a quantum state in Sect.~\ref{sub:concrete_state}
The importance of the positivity of $\rho$ in the is that it ensures the positive definiteness of the \emph{energy inner product} on the space of solutions
\begin{align}
  \langle A \,|\, B \rangle_{\mathrm{en}} &\defn \frac{1}{2} \int_\Sigma \big( \dif \conj{A} \wedge (\iprod{N} \conlaw \dif B) - \conlaw \dif \conj{A} \wedge (\iprod{N} \dif B) \big) \label{eq:energy_prod}\\
  &\mathrel{\phantom{\defn}\mathllap{=}} \frac{1}{2} \int_\Sigma \big( \conj{A} \wedge \conlaw \dif \lie_N B - \conlaw \dif \conj{A} \wedge \lie_N B \big).\label{eq:energy_prod2}
\end{align}
The two equivalent formulations correspond to each other via Stokes' theorem\footnote{For this relation one should assume that $\Sigma$ has no boundary. At the very least one must require that the boundary of~$\Sigma$ does not intersect with the support of $A$ and $B$.} and Cartan's magic formula which relates the Lie derivative, the exterior derivative and the interior product.

\emph{The energy inner product is positive definite and Hermitian.}
Hermiticity can be seen from the fact that
\begin{align*}
  \iprod{N} (\conlaw \dif \conj{A} \wedge \dif B) &= \iprod{N} (\dif \conj{A} \wedge \conlaw \dif B)
\intertext{implies that}
  \conlaw \dif \conj{A} \wedge (\iprod{N} \dif B) - \dif \conj{A} \wedge (\iprod{N} \conlaw \dif B) &= \conlaw \dif B \wedge (\iprod{N} \dif \conj{A}) - \dif B \wedge (\iprod{N} \conlaw \dif \conj{A}),
\end{align*}
\ie, that the integrand of~\eqref{eq:energy_prod} is pointwise Hermitian.
Positivity is clear from
\begin{align*}
  \langle A \,|\, A \rangle_{\mathrm{en}} = \int_\Sigma T_N(A) = \int_M n \wedge T_N(A) = \int_M \rho > 0
\end{align*}
for $A$ that are not pure gauge.
Furthermore, observe that by~\eqref{eq:energy_prod2} \emph{the energy inner product is closely related to the symplectic form}~\eqref{eq:sympl2} by direct comparison of the corresponding expressions
\begin{equation*}
  \langle A \,|\, B \rangle_{\mathrm{en}} = \varsigma^\Gamma(\conj{A}, \lie_N B)
\end{equation*}
if $n \in \Gamma$ is a hyperbolicity covector, $\Sigma$ a $\Gamma$-Cauchy surface with $\ker n = T\Sigma$ and $N$ is a generalized Killing vector field dual to $n$ (\ie, $n(N) = 1$) such that~\eqref{eq:positive_constlaw} is satisfied.
This relationship demonstrates, as a consequence of the independence of $\varsigma^\Gamma$ on the choice of the $\Gamma$-Cauchy surface~$\Sigma$, see Sect.~\ref{sub:sympl}, that also $\langle A \,|\, B \rangle_{\mathrm{en}}$ is independent of this choice.

Thus, with help of the kinematic energy-momentum of the theory, we found a way to construct a positive inner product on the space of solutions of the homogeneous field equations for a certain class of constitutive laws.
This is the class of theories of electrodynamics which we consider now for quantization.

\section{Quantum field theory}\label{sec:quant}

Henceforth we shall only discuss theories with one hyperbolicity double cone for which the energy inner product is positive; some reasons for this were already discussed in Sect.~\ref{sub:sympl} and $\ref{sub:energydensity}$ and other reasons will become clear in Sect.~\ref{sub:norder} and \ref{sub:concrete_state}.
Consequently we choose here a preferred hyperbolicity cone $\Gamma$ and drop the $\Gamma$ sub- and superscripts as no confusion can arise.

Moreover, we will restrict to the case, where the principal symbol $M$ given in~\eqref{eq:principal} can be considered to be of real principal type.
What we mean by this is explained in App.~\ref{app:propsing}.
Essentially we will require that $\chi$ is given either by a Lorentzian metric or $\mathcal{G}(k, k, k, \cdot) \neq 0$ for all $k$ such that $\mathcal{G}(k) = 0$.
Note that this assumption is related to the concept of bihyperbolicity of the principal symbol introduced in~\cite{Raetzel:2010je}.

\subsection{Algebraic quantization}\label{sub:alg}

In this section we will quantize the phase space $\mathfrak{S}^*_{\mathrm{sc}}$ introduced in Sect.~\ref{sub:sympl} using the algebraic approach.
We will follow roughly the general approach of~\cite{dimock:1980} which has been
quite successful in quantum field theory on curved spacetimes.

Denote by $\mathfrak{A}$ the unital ${}^*$-algebra finitely generated by the \emph{quantum field} $\mathcal{A} : \Omega^3_{c,\dif}(M) \to \mathfrak{A}$ with the properties\footnote{We denote the complex conjugate of $z$ by $\conj{z}$.}
\begin{description}[labelwidth=8em,align=right]
  \item[Linearity] $\mathcal{A}(\alpha J + \beta K) = \alpha \mathcal{A}(J) + \beta \mathcal{A}(K)$ for all $\alpha, \beta \in \CC$,
  \item[Hermicity] $\mathcal{A}(J)^* = \mathcal{A}(\conj{J})$,
  \item[Field equation] $\mathcal{A}(P A) = 0$,
  \item[CCR] $[\mathcal{A}(J), \mathcal{A}(K)] = \im \sigma(J, K) \one$,
\end{description}
for all $J, K \in \Omega^3_{c,\dif}(M)$ and $A \in \Omega^1_c(M)$; we denote the unit element of $\mathfrak{A}$ by $\one$.
\emph{In words, the quantum field is linear, its adjoint is given by complex conjugation of its argument, it is a weak solution of the field equation and it implements the canonical commutation relations (CCR) given by the (pre-)symplectic form $\sigma$.}
Observe that Einstein causality holds, \viz, $\mathcal{A}$ smeared with spacelike related $3$-forms $J, K$ commute, because of the support properties of $\sigma$.
We call the algebra $\mathfrak{A}$ the \emph{field algebra of pre-metric electrodynamics}.
To give an example, a typical element of $\mathfrak{A}$ is
\begin{equation*}
  \mathcal{A}(J_1) + \mathcal{A}(J_{21}) \mathcal{A}(J_{22}) + \mathcal{A}(J_{31}) \mathcal{A}(J_{32}) \mathcal{A}(J_{33}) + \dotsb
\end{equation*}
with finitely many terms.

Sometimes it is useful to consider the completion $\mathfrak{A}^{\mathrm{cpl}}$ of $\mathfrak{A}$ in its natural%
\footnote{\label{note:topology}%
  The `natural' topology of $\mathfrak{A}$ is that induced (via the direct sum, quotient and subspace topology) by the test function topology on $\Omega^3_c(M)$.
  This uses the fact that the field algebra is the quotient of the tensor algebra $\bigoplus_n \mathfrak{S}_{\mathrm{sc}}^{\otimes n}$ by the commutation relations.
  Also note that, in the test function topology, $\Omega^3_c(M)$ is a nuclear Fréchet space so that the usual notions of tensor products coincide and a Schwartz kernel theorem can be formulated.
}
topology.
Consider the continuous extension of $\mathcal{A}^{\otimes n}$ to the map (denoted by the same symbol)
\begin{equation}\label{eq:Acompl}
  \mathcal{A}^{\otimes n} : \big(\Omega^3_{c,\dif}(M)^{\otimes n}\big)^{\mathrm{cpl}} \to \mathfrak{A}^{\mathrm{cpl}}.
\end{equation}
To get a better idea of this map, we may write formally
\begin{equation*}
  \mathcal{A}^{\otimes n}(J) = \int_{M^{\times n}} \big( \mathcal{A}(x_1) \otimes \dotsm \otimes \mathcal{A}(x_n) \big)\, J(x_1, \dotsc, x_n).
\end{equation*}
The maps $\mathcal{A}^{\otimes n}$ can be used to generate the more general elements in the completion $\mathfrak{A}^{\mathrm{cpl}}$.

We remark that the quantum field $\mathcal{A}$ can be understood as a ${}^*$-algebra-valued distribution on $\Omega^3_{c,\dif}(M)$; we already used this fact in the previous equation.
This is quite similar to the usual situation in `non-algebraic' quantum field theory, where the quantum field can be rigorously understood as an operator-valued distribution.
Nevertheless, this similarity should be handled with care as $\mathfrak{A}$ is no Hilbert space.
In the next section this similarity will become clearer after we introduced the notions of states and the famous GNS theorem.

Furthermore, we remark that the effect of taking the quotient by the canonical commutation relations (CCR) is essentially that of modifying the product in the algebra $\mathfrak{A}$.
An approach which makes this observation concrete is that of deformation quantization, see \eg~\cite{brunetti:2009a}.
The deformation quantization approach is very useful in perturbative algebraic quantum field theory, a subject that we will not discuss any further here.
We mention, however, that some of the notions of Sect.~\ref{sub:norder} can be made more precise and general using techniques from deformation quantization.

\subsection{Quantum states}\label{sub:states}

While the algebra constructed in the previous section, gives an abstract mathematical description of `observables', \ie, operations performed on a physical system, the concept of states gives an abstract mathematical description of the preparation of the physical system.
Then, observables act upon this prepared system.
The abstract discussion of states is often avoided in QFT on Minkowski spacetime because there is one preferred state, the Poincaré-invariant vacuum state.
More general spacetimes possess no symmetries and no construction for preferred (ground) states exists.
We are working with pre-metric electrodynamics on $M=\RR^4$ and the field equations have constant coefficients.
In this situation we can work just like in QFT on Minkowski spacetime and attempt to construct translation-invariant states.
Thus we could, in principle, avoid the general discussion below.
However, for conceptual clarity and also as preparation for an eventual construction of states on non-static backgrounds (\ie, position- and time-dependent constitutive laws), we will give a general but concise discussion of quantum states on the field algebra $\mathfrak{A}$.
Later, after having introduced the microlocal spectrum condition in Sect.~\ref{sub:norder}, we will give a concrete construction of a quantum state in Sect.~\ref{sub:concrete_state}.

States on $\mathfrak{A}$ (and equivalently on $\mathfrak{A}^{\mathrm{cpl}}$) are the normalized positive elements of $\mathfrak{A}'$, the topological\footref{note:topology} dual of $\mathfrak{A}$.
That means, $\omega \in \mathfrak{A}'$ (\ie, $\omega : \mathfrak{A} \to \CC$ is linear and continuous) is a state on the field algebra $\mathfrak{A}$ if
\begin{description}[labelwidth=8em,align=right]
  \item[Normalization] $\omega(\mathbf{1}) = 1\;$ and
  \item[Positivity] $\omega(a^* a) \geq 0\;$ for all $a \in \mathfrak{A}$.
\end{description}
Each state $\omega \in \mathfrak{A}'$ can be represented by a hierarchy of \emph{$n$-point distributions} $(\omega_n)_{n \geq 0}$ with $\omega_n \in (\Omega^3_c(M)^{\otimes n})'$, \ie, each $\omega_n : \Omega^3_c(M)^{\otimes n} \to \CC$ is multilinear and continuous, by setting
\begin{equation*}
  \omega_n(J_1, \dotsc, J_n) \defn \omega\big(\mathcal{A}(J_1) \,\dotsm\, \mathcal{A}(J_n)\big) = \omega\big(\mathcal{A}^{\otimes n}(J_1 \otimes \dotsm \otimes J_n)\big).
\end{equation*}
Clearly, each $\omega_n$ can be continuously extended to $(\Omega^3_c(M)^{\otimes n})^{\mathrm{cpl}}$ so that we may equivalently define the $n$-point distributions by $\omega_n(J) = \omega(\mathcal{A}^{\otimes n}(J))$.

It follows from the properties of the quantum field $\mathcal{A}$, that an admissible $n$-point distribution~$\omega_n$ must be a weak solution of the field equation in each argument
\begin{equation*}
  \omega_n(J_1, \dotsc, J_{i-1}, P A, J_{i+1}, \dotsc, J_n) = 0
\end{equation*}
and satisfy (weakly) the commutation relation given by the symplectic form $\sigma$
\begin{multline*}
  \omega_n(J_1, \dotsc, J_i, J_{i+1}, \dotsc, J_n) - \omega_n(J_1, \dotsc, J_{i+1}, J_i, \dotsc, J_n) \\= \im \sigma(J_i, J_{i+1}) \omega_{n-2}(J_1, \dotsc, J_{i-1}, J_{i+2}, \dotsc, J_n),
\end{multline*}
for all $J_i \in \Omega^3_{c,\dif}(M)$ and $A \in \Omega^1_c(M)$.
This representation in terms of distributions is non-unique as two distinct $\omega_n$ and $\omega'_n$ are gauge-equivalent if
\begin{equation*}
  \omega_n(J_1, \dotsc, J_n) = \omega'_n(J_1, \dotsc, J_n)
\end{equation*}
for all closed $3$-forms $J_i$.
In other words, \emph{there is a gauge freedom in fixing $\omega_n$}.
This is exactly the same gauge freedom that we encountered when we constructed the inverse of the field equations in Sect.~\ref{sub:inv}.
We can see this by considering the bidistribution defined by $\int \cdot \wedge (E^\Gamma \,\cdot\,)$, which is independent of the gauge of $E^\Gamma$ when smeared with conserved $3$-forms.

In some publications concerned with states for the electromagnetic vector potential, \eg~\cite{fewster:2003} or~\cite{dappiaggi:2013} by one of the authors, it is actually claimed that $\omega(\mathcal{A}(J_1), \dotsc, \mathcal{A}(J_n))$ do not define distributions because $J_i$ are required to be conserved.
The discussion above makes this statement more precise.
Namely, \emph{a state defines a hierarchy of (gauge-)equivalence classes of distributions}.

One often restricts to the class of \emph{quasi-free states}%
\footnote{%
  Quasi-free states are the natural states in free theories, which is evidently the case here, see \eg\ the quadratic action~\eqref{eq:action}.
}
(also called \emph{Gaussian states}).
These states are completely characterized by their two-point distribution so that all even $n$-point distributions are given by
\begin{equation*}
  \omega_n(J_1, \dotsc, J_n)
  = \sum_{\sigma} \omega_2(J_{\sigma(1)}, J_{\sigma(2)}) \,\dotsm\, \omega_2(J_{\sigma(n-1)}, J_{\sigma(n)}),
\end{equation*}
where the sum is over all ordered pairings, \ie, over all permutations $\sigma$ of $\{1, \dotsc, n\}$ such that $\sigma(1) < \sigma(3) < \dotsb < \sigma(n-1)$ and $\sigma(1) < \sigma(2), \dotsc, \sigma(n-1) < \sigma(n)$, and all odd $n$-point distributions vanish.
Often one does not distinguish between a quasi-free state $\omega$ and its two-point distribution $\omega_2$.
Let us emphasize that a two-point distributions $\omega_2$ is an element of $(\Omega^3_c(M)^{\otimes 2})'$ and satisfies the properties
\begin{equation}\label{eq:two_point_prop}
  \omega_2(\conj{J}, J) \geq 0,
  \quad
  \omega_2(J, P A) = 0 = \omega_2(P A, J)
  \quad\text{and}\quad
  \omega_2(J, K) - \omega_2(K, J) = \im \sigma(J, K)
\end{equation}
for all $J, K \in \Omega^3_{c,\dif}(M)$ and $A \in \Omega^1_c(M)$.

We remark that, once a state has been fixed, one can work again in the familiar setting of Hilbert spaces.
The transition from the ${}^*$-algebra $\mathfrak{A}$ and a state $\omega$ is achieved by the GNS theorem, see \eg~\cite{powers:1971}.
It states that a state on a ${}^*$-algebra induces a representation of the algebra on a Hilbert space with a cyclic (``vacuum'') vector.
Noticing that a state induces a positive but (possibly) degenerate inner product on $\mathfrak{A}$, this theorem is essentially proved by quotienting through the null space and then completing the resulting pre-Hilbert space.
One can then see that quasi-free states correspond in this way to Fock spaces and then the quantum field can be represented in terms of creation and annihilation operators, see \eg~\cite[Chap.~4]{wald:1994}.

\subsection{Normal ordering and the (microlocal) spectrum condition}\label{sub:norder}

In quantum field theory one often encounters products of quantum fields at a point.
Such objects cannot be described by elements of the field algebra $\mathcal{A}$ or $\mathcal{A}^{\mathrm{cpl}}$; it is necessary to enlarge this algebra.
On the other hand, the space of states discussed in the previous section certainly contains many unphysical states.

In the following we will argue constructively and sometimes formally to derive conditions that mathematically well-behaved states and normal ordering prescriptions must satisfy.
As it turns out, these conditions are also physically desirable.

A \emph{normal ordering} (or \emph{Wick ordering}) prescription $\norder{\,\cdot\,}$ with respect to a bidistribution $\lambda_2 \in (\Omega^3_c(M)^{\otimes 2})'$ can be (formally) implemented recursively by~\cite{brunetti:1996}
\begin{align*}
  \norder{\mathcal{A}}(x) &= \mathcal{A}(x) \\
  \norder{\mathcal{A}^{\otimes (n+1)}}(x_1, \dotsc, x_{n+1}) &= \norder{\mathcal{A}^{\otimes n}}(x_1, \dotsc, x_n) \mathcal{A}(x_{n+1}) \\&\quad - \sum_{i=1}^n \norder{\mathcal{A}^{\otimes (n-1)}}(x_1, \dotsc, x_{i-1}, x_{i+1}, \dotsc, x_n) \lambda_2(x_i, x_{n+1}).
\end{align*}
Analogously to~\eqref{eq:Acompl}, we can extend the definition of the normal ordered fields to maps from $(\Omega^3_{c,\dif}(M))^{\mathrm{cpl}}$.

We remark that if $\lambda_2 = \omega_2$ is the two-point distribution of a quasi-free state, then the normal ordering defined above is equivalent to the usual normal ordering of creation and annihilation operators in the Fock space representation of the state $\omega$.
Moreover, observe that in this case all $\norder{\mathcal{A}^{\otimes n}}$ are symmetric.
For now we will not require that $\lambda_2$ satisfies the properties~\eqref{eq:two_point_prop} of a two-point distribution of a state, although below we will see that $\lambda_2$ should be equal to $\omega_2$ up to a smooth remainder so that these properties are also satisfied up to a smooth remainder.
Such a more general choice comes with advantages (\eg, there might exist a natural candidate for $\lambda_2$ but no natural candidate for $\omega_2$) and disadvantages (\eg, if $\lambda_2$ is not a weak solution of the field equations, the normal ordered field are also not weak solutions).

In the remainder of this section we will motivate a physical and mathematical requirement on the two-point distribution $\omega_2$ and also on $\lambda_2$.
For this purpose we will use the concept of the wavefront set of a distribution; we refer to~\cite{brouder:2014} for an introduction to the wavefront set including several examples.
To formulate this requirement in a convenient way, let
\begin{equation*}
  \mathcal{N} \defn \big\{ (x, k) \in T^*\!M \setminus \{0\} \;\big|\; \mathcal{G}(x, k) = 0 \big\}
\end{equation*}
be the zero set of $\mathcal{G}(x, k)$ as a function on the cotangent bundle $T^*M$.
In other words, it is the characteristic set of $\mathcal{G}(x, \partial)$.
Moreover, we decompose $\mathcal{N}$ into two disconnected components $\mathcal{N} = \mathcal{N}^+ \cup \mathcal{N}^-$, where
\begin{equation*}
  \mathcal{N}^\pm \defn \big\{ (x, k) \in \mathcal{N} \;\big|\; \mathcal{G}(x; n, n, n, k) < 0, n \in \pm \Gamma \big\}.
\end{equation*}
In words: $\mathcal{N}^\pm$ are the future (+) and past (-) pointing null-momenta with respect to the time-orientation induced by the hyperbolicty cone $\Gamma$.
Physically these are interpreted as null-momenta with positive or negative energy.
This decomposition of null-momenta in positive and negative energy with respect to a hyperbolicity covector is certainly always possible for constitutive densities whose principal symbol is of real principal type and is a special case of the discussion of energy-distinguishing dispersion relations in~\cite{Raetzel:2010je}.

We will show that a good choice of two-point distributions should satisfy the so-called \emph{microlocal spectrum condition}
\begin{equation}\label{eq:microlocal}
  \WF(\omega_2) = \WF(\lambda_2) = \big\{ (x_1, k_1; x_2, -k_2) \in \mathcal{N}^+ \times \mathcal{N}^- \;\big|\; (x_1, k_1) \sim (x_2, k_2) \big\},
\end{equation}
where $(x_1, k_1) \sim (x_2, k_2)$ means that $(x_1, k_1)$ and $(x_2, k_2)$ lie in the same orbit of the Hamiltonian flow $X_P$ of $P$.
In most cases we have
\begin{equation}
  X_P(x, k) = \frac{\partial \mathcal{G}(x, k)}{\partial k_a} \frac{\partial}{\partial x^a} - \frac{\partial \mathcal{G}(x, k)}{\partial x^a} \frac{\partial}{\partial k_a}
\end{equation}
although there are some subleties when $X_P$ is vanishing on $\mathcal{N}$, see App.~\ref{app:propsing} for details.
For example, for a constant constitutive law this means $k_1 = k = -k_2$ with $k \in \mathcal{N}^+$ and $x_1, x_2$ are connected by a curve whose tangent vector field is given by $\mathcal{G}^{abcd} k_b k_c k_d$.
Observe that our microlocal spectrum condition is an obvious generalization of an equivalent condition from quantum field theory on curved spacetimes~\cite{brunetti:1996,radzikowski:1996}.
Roughly speaking, the microlocal spectrum condition says that particles with sufficiently large momenta have a positive energy.

Applying normal ordering to a product of two fields, we obtain
\begin{equation}\label{eq:norder_square}
  \norder{\mathcal{A}^{\otimes 2}}(x_1, x_2)
  = A(x_1) A(x_2) - \lambda_2(x_1, x_2) \one
  = \mathcal{A}^{\otimes 2}(x_1, x_2) - \lambda_2(x_1, x_2) \one.
\end{equation}
An important quantity in quantum field theory is the Wick square at a point
\begin{equation*}
  \omega\big(\norder{\mathcal{A}^{\otimes 2}}(x)\big) = (\omega_2 - \lambda_2)(x, x),
\end{equation*}
\ie, the coincidence limit of~\eqref{eq:norder_square} evaluated in a state $\omega$; it is related to the energy density of the quantum field in the state $\omega$.
For this expression and all of its derivatives to be well-defined we will require that $\omega_2 - \lambda_2$ is smooth.
That is, we demand
\begin{equation}\label{eq:WF_assumption1}
  \WF(\lambda_2) = \WF(\omega_2),
\end{equation}
\viz, the wavefront sets of $\lambda_2$ and $\omega_2$ shall agree.

Applying normal ordering to products of more than three fields $\mathcal{A}$, we see that tensors products of $\lambda_2$ appear, \eg, $\lambda_2^{\;\otimes 2}(x_1, x_2, x_3, x_4)$.
Although such tensor products are well-defined distributions, we have to be more careful if we wish to smear them with distributions.
Indeed, $J$ \emph{must be smooth} if the singular directions of $\lambda_2$ at each point in $M \times M$ are \emph{not} contained in a cone $\Gamma$ such that $\Gamma \cap -\Gamma = \emptyset$.
This may be seen by the fact that only then powers of $\lambda_2$ are well-defined distributions.
Therefore we demand that
\begin{equation}\label{eq:WF_assumption2}
  \WF(\omega_2) \cap -\WF(\omega_2) = \emptyset,
\end{equation}
where we set $(x_1, k_1; x_2, -k_2) \in \WF(\omega_2) \Leftrightarrow: (x_1, -k_1; x_2, k_2) \in -\WF(\omega_2)$.
We remark that this is not merely a wish for mathematical convenience because physically relevant quantities, \eg, the fluctuations of the energy density in a state, make sense only if powers of the two-point function are well-defined distributions.

We will now investigate the microlocal consequences of~\eqref{eq:WF_assumption1} and~\eqref{eq:WF_assumption2}.
Being a solution in both of its arguments, it follows from~\cite[Thm.~18.1.28]{Hoermander3} that the wavefront set of $\omega_2$ is included in the characteristic set of the operators $P \otimes \id$ and $\id \otimes P$ so that
\begin{equation*}
  \WF(\omega_2) \subset \mathcal{N} \times \mathcal{N}.
\end{equation*}
In fact, by propagation of singularities~\cite[Thm.~26.1.1]{Hoermander4} and App.~\ref{app:propsing}, we see that
\begin{equation*}
  (x_1, k_1; x_2, k_2) \in \WF(\omega_2) \implies (x_1', k_1'; x_2', k_2') \in \WF(\omega_2)
\end{equation*}
for all $(x_1', k_1') \sim (x_1, k_1)$ and $(x_2', k_2') \sim (x_2, k_2)$.
Note now that
\begin{equation*}
  \WF(\sigma) = \WF(\upDelta) = \big\{ (x_1, k_1; x_2, -k_2) \in \mathcal{N} \times \mathcal{N} \;\big|\; (x_1, k_1) \sim (x_2, k_2) \big\},
\end{equation*}
which is a consequence of the propagation of singularities, see \eg~\cite[Sect.~6.5]{Duistermaat}.
Since the antisymmetric part of $\omega_2$ is given by $\im \sigma$, see~\eqref{eq:two_point_prop}, each $(x_1, k_1; x_2, -k_2) \in \mathcal{N} \times \mathcal{N}$ is included either in $\WF(\omega_2)$ or in $-\WF(\omega_2)$ but by~\eqref{eq:WF_assumption2} not in both.
In other words, at each $(x_1, x_2) \in M \times M$ the singular directions of $\omega_2$ are included either in $\mathcal{N}^+_{x_1} \times \mathcal{N}^-_{x_2}$ or $\mathcal{N}^-_{x_1} \times \mathcal{N}^+_{x_2}$.
Next we use that any two points $x_1$ and $x_2$ can be connected via the Hamiltonian flow $X_P$, \viz, there exist $x_3$ and $k_1, k_2, k_3, k_3' \in \mathcal{N}$ such that $(x_1, k_1) \sim (x_3, k_3)$ and $(x_2, k_2) \sim (x_3, k_3')$.
It follows that the singular directions at some point in $M \times M$ imply the singular directions at all other points:
\begin{equation*}
  \WF(\omega_2) \subset \begin{cases}
    \mathcal{N}^+ \times \mathcal{N}_-, \quad\text{or} \\
    \mathcal{N}^- \times \mathcal{N}_+.
  \end{cases}
\end{equation*}
Therefore, $\WF(\omega_2) \cap \WF({}^t\!\omega_2) = \emptyset$ yet $\WF(\omega_2) \cup \WF({}^t\!\omega_2) = \WF(\sigma)$, where ${}^t\!\omega_2$ is the transpose of $\omega_2$ in the sense of bilinear maps, \viz, the arguments are exchanged.
We finally conclude that the only possibilities are
\begin{equation*}
  \WF(\omega_2) = \begin{cases}
    \big\{ (x_1, k_1; x_2, -k_2) \in \mathcal{N}^+ \times \mathcal{N}^- \;\big|\; (x_1, k_1) \sim (x_2, k_2) \big\}, \\
    \big\{ (x_1, k_1; x_2, -k_2) \in \mathcal{N}^- \times \mathcal{N}^+ \;\big|\; (x_1, k_1) \sim (x_2, k_2) \big\}.
  \end{cases}
\end{equation*}
The choice between the two is just a convention and is related to the choice of the sign in the Fourier transform.
It is customary to choose the first possibility and we will do the same.
Henceforth we will require that $\lambda_2$ and $\omega_2$ satisfy the microlocal spectrum condition~\eqref{eq:microlocal}.

\emph{Consequently, at each point $x$, the singular directions of $\omega_2$ and $\lambda_2$ are contained in the cone $\mathcal{N}^+_x \times \mathcal{N}^-_x$.
Therefore they can be smeared with compactly supported, distributional $3$-forms $J \in (\Omega^3(M)^{\otimes n})'$ that satisfy $\WF(J) \cap \mathcal{N}^{\times n} = \emptyset$.}
Here it is beneficial to restrict to $J$ that are symmetric under exchange of arguments so that the symmetry of the normal ordered fields is enforced even when $\lambda_2$ is not the two-point distribution of a state.
The spaces of such $J$ form the basis for the construction of the so-called \emph{algebra of Wick polynomials}.
As we only wanted to motivate the microlocal spectrum condition, we will not continue to construct this algebra.
The interested reader should have no difficulty to complete the construction using \eg~\cite{brunetti:1996,hollands:2001} as references.

\subsection{Quantum states for constant constitutive laws}\label{sub:concrete_state}

In this section we will outline a construction of states that satisfy the microlocal spectrum condtion~\eqref{eq:microlocal} in the case of a constant constitutive law which has a positive energy density~\eqref{eq:positive_constlaw} with respect to the vector field
\begin{equation*}
  N^a = \frac{\mathcal{G}^{abcd} n_b n_c n_d}{\mathcal{G}(n, n, n, n)}
\end{equation*}
for some constant hyperbolicity vector $n \in \Gamma$.
Clearly $N$ is a generalized Killing vector field because the constitutive density and $N$ are both constant.

To do so we use the energy inner product on the space of solutions of the homogeneous field equations $\mathfrak{S}_{\mathrm{sc}}$ defined in~\eqref{eq:energy_prod2} to show that the kernel of the Lie derivative $\lie_N$ on $\mathfrak{S}_{\mathrm{sc}}$ is trivial.
This follows immediately from the definition of the energy inner product because
\begin{equation*}
  \lie_N A = 0
  \quad\Rightarrow\quad
  \langle A \,|\, A \rangle_{\mathrm{en}} = \frac12 \varsigma(\conj{A}, \lie_N A) = 0
  \quad\Rightarrow\quad
  A = 0.
\end{equation*}

Completing $\mathfrak{S}_{\mathrm{sc}}$ with respect to the energy product, we obtain a Hilbert space $\mathcal{H}_{\mathrm{en}}$.
Clearly, $\lie_N$ is densely defined on $\mathcal{H}_{\mathrm{en}}$ and it is closeable because it is anti-Hermitian; we denote its closure by the same symbol.
We perform a polar decomposition of $\lie_N$ to define
\begin{equation*}
  U\, \abs{\lie_N} \defn \lie_N.
\end{equation*}
Since $\lie_N$ has a trivial kernel on the solution space, also $\abs{\lie_N}$ has a trivial kernel there and we can define
\begin{equation*}
  \mu(A, B) \defn \langle \conj{A} \,|\, \abs{\lie_N}^{-1} B \rangle_{\mathrm{en}} = \frac{1}{2} \varsigma(A, U B)
\end{equation*}
for $A, B \in \mathfrak{S}_{\mathrm{sc}}$.
We remark that $\mu$ is symmetric because $\abs{\lie_N}$ is self-adjoint or, equivalently, because $U$ is an anti-involution (also called complex structure) that tames $\varsigma$.
Setting
\begin{equation}\label{eq:2ptstate}
  \omega_2(J, K) \defn \mu(\upDelta J, \upDelta K) + \frac{\im}{2} \sigma(J, K),
\end{equation}
we have thus defined the \emph{two-point distribution of a pure quasi-free state}.

The state defined by \emph{$\omega_2$ is a ground state} with respect to the symmetry given by the Killing vector field $N$.
Namely, it satisfies
\begin{equation}\label{eq:ground_state}
  -\im \omega_2(\conj{J}, \lie_N J) \geq 0
\end{equation}
as $\upDelta$ commutes with $\lie_N$.
Note that $\lie_N$ is simply the generator of translations in the direction of~$N$.
Denote by $\tau_t$ the pullback by the flow generated by $N$; this is the usual translation map along~$N$, \eg, $\tau_t f(s, \vec{x}) = f(s - t, \vec{x})$ for a function $f$ on $M$.
Then one can show, \cf~\cite[App.~1]{Sahlmann}, that the condition~\eqref{eq:ground_state} is equivalent to
\begin{equation}\label{eq:ground_state2}
  \int_\RR f(t)\, \omega_2(J, \tau_t K)\, \dif t = 0
\end{equation}
for functions $f$ such that its Fourier transform is compactly supported on the negative half-line $(-\infty, 0)$.

We will now investigate the microlocal properties of $\omega_2$, \ie, whether $\omega_2$ satisfies the microlocal spectrum condition~\eqref{eq:microlocal}.
By construction, $\omega_2$ is a solution of the field equation with anti-symmetric part given by the Pauli--Jordan propagator and thus it follows that its wavefront set satisfies
\begin{equation*}
  \WF(\omega_2) \subset \big\{ (x_1, k_1; x_2, -k_2) \in \mathcal{N} \times \mathcal{N} \;\big|\; (x_1, k_1) \sim (x_2, k_2) \big\}.
\end{equation*}
Following the arguments of the previous section, we now only need to show that $\WF(\omega_2) \subset \mathcal{N}^+ \times \mathcal{N}^-$ holds.

As \eg\ observed in~\cite{Verch}, the wavefront set of a distribution is closely related to the spectral properties of the action of the translation map on it.
Among other things, this fact was used in~\cite{Sahlmann} to show that ground states for QFT on curved spacetime satisfy the usual metric-based microlocal spectrum condition.
However, it is clear that their proof generalizes straightforwardly also to our case.

We give a sketch of the argument:
Let $h \in C^\infty(\RR)$ have a compactly supported Fourier transform.
Then
\begin{equation*}
  \lim_{k \to -\infty} \int_\RR h(t)\, \e^{-\im k t}\, \omega_2(J, \tau_t K)\, \dif t = \int_\RR f_k(t)\, \omega_2(J, \tau_t K)\, \dif t = 0
\end{equation*}
because for sufficiently large negative $k$ the Fourier transform $\what{f_k}(p) = \what{h}(p + k)$ is supported in $(-\infty, 0)$.
One can then deduce that the wavefront set for the second argument of $\omega_2$ must lie in~$\mathcal{N}^-$ and thus \emph{the microlocal spectrum condition~\eqref{eq:microlocal} holds}.

\section{Discussion}

In this article we covariantly quantized pre-metric electrodynamics with constant coefficient constitutive density.
The first important result towards this goal was the explicit construction of the quasi-inverse of the principal symbol of the field equations, \ie, the `Fourier representation' of the photon propagator in pre-metric electrodynamics, in Sect.~\ref{sub:quasiinv}.
The idea to construct the quasi-inverse is the same as that of Itin~\cite{Itin:2007aa,Itin:2015tdi} by using the second adjoint.\footnote{Note that~\cite{Itin:2007aa} contains a small mistake due to a missing anti-symmetrization which is corrected in~\cite{Itin:2015tdi} and, independently, here.}
However, our derivation emphasizes more the role of the gauge freedom and its fixing.
Thereby we obtain a precise characterization and parametrization of the gauge freedom of the theory.

Prerequisites for the application of locally covariant quantization are that the theory defining constitutive density leads to a hyperbolic Fresnel polynomial and yields a positive energy-momentum inner product.

The hyperbolic Fresnel polynomial is essential for the causal behaviour of the theory, as described in Sect.~\ref{sub:causality}.
We introduced the physically important notions like the causal future and the causal past of subsets of spacetime as well as the notion of Cauchy surfaces in the context of pre-metric electrodynamics.
These notions allowed us to discuss the causal behaviour of the solutions to the field equations of pre-metric electrodynamics in Sects.~\ref{sub:inv} and~\ref{sub:causp}.
There we used the same language as in the study of solutions to the metric wave equation on Lorentzian spacetimes.

The positive energy-momentum inner product defined in~\eqref{eq:energy_prod} ensures the existence of a ground state on the field algebra, which we constructed explicitly in~\eqref{eq:2ptstate}.
With the identification of the relation between the positivity of the energy-momentum inner product and properties of the constitutive density in~\eqref{eq:positive_constlaw}, we systematically connect the results of an earlier canonical approach to the quantization of pre-metric electrodynamics~\cite{Rivera:2011rx}, the construction of quantum states on static spacetimes, see \eg~\cite{wald:1994}, and the axiomatic approach to electrodynamics~\cite{Hehl}.
It is likely that the prerequisites for the constitutive law we demand here are more restrictive than the requirement of bihyperbolicity and the energy-distinguishing property in~\cite{Raetzel:2010je}.

Constitutive densities which satisfy the requirements just mentioned define linear theories of electrodynamics which are as well behaved as Maxwell electrodynamics based on a spacetime metric.
Thus those constitutive densities serve equally well as (geometric) background to define the field equations of a physical field theory as a Lorentzian spacetime metric.

With the local covariant quantization of pre-metric electrodynamics with constant constitutive law we lay the foundation for the quantization of the general case with non-constant constitutive law.
Analogue to the extension of quantum field theory on Minkowski spacetime to quantum field theory on generally curved spacetime, the methods of algebraic quantum field theory, which we already used in this paper, are suitable to extend the construction.
A main future task in the general case is the construction of advanced and retarded propagators for the field equations.

The mathematical rigorous framework we used here allows a direct analysis of quantum effects, such as the Casimir effect or quantum energy inequalities, on the basis of the constitutive density as geometric background field.
Due to this more complex background structure compared to a metric geometry, we expect qualitative and quantitative deviations to the known results.

A large field of concrete applications for pre-metric electrodynamics is the description of electrodynamics in media.
Systems which are suitable to perform explicit calculations of the quantum effects are linear permeable media and in particular birefringent uniaxial crystals, as they were discussed as examples already in this article.
These applications continue the project of a locally covariant quantum field theory point of view on the results which were obtained on the derivation of the Casimir Effect in birefringent optical media by applying canonical quantization to pre-metric electrodynamics in~\cite{Rivera:2011rx}.

With this paper we continued to demonstrate that the wave equation on Lorentzian spacetime is by far not the only equation of interest and an immediate open question is if one can realize field equations for scalar fields and spinors whose solutions follow the same causal structure as the vector potential in pre-metric electrodynamics.
For the scalar field one may investigate the Fresnel partial differential equation~\eqref{eq:Gdiffeq}, while a Dirac equation may be constructed by considering a first order equation which is consistent with the Fresnel partial differential equation.
Having included the description of spinors on the background geometry defined by the constitutive density one may even aim on a complete formulation of Quantum General Linear Electrodynamics.

\begin{acknowledgments}
  We would like to thank Claudio Dappiaggi, Jan Dereziński, Thomas-Paul Hack, Robin Tucker and Volker Perlick for useful discussions.
  We also thank Igor Khavkine for comments on an earlier version of this article.
  We are grateful for the kind hospitality extended by the ZARM at the University of Bremen where this work was initiated.
  During his time in Bremen, D.S. was supported through a Riemann fellowship of the Riemann Center for Geometry and Physics, University of Hannover.
  The work of D.S. was supported by the National Science Center (NCN) based on the decision No. DEC-2015/16/S/ST1/00473.
\end{acknowledgments}

\appendix

\section{Uniaxial crystals}\label{app:Uniaxial}

As an example beyond Maxwell electrodynamics we mentioned the uniaxial crystal in Sects.~\ref{sub:field_eq} and~\ref{sub:quasiinv}.
In this appendix we discuss the derivation of the constitutive law~\eqref{eq:condensuni}, the Fresnel polynomial~\eqref{eq:fresneluni}, the $Q$ matrix~\eqref{eq:uniaxial_Q} and possible gauge choices in the uniaxial crystal.

Uniaxial crystals are simple media in which birefringence occurs.
They can be described in terms of linear dielectric media with an dielectricity $\epsilon$ which has two distinguished eigenvalues and a trivial (magnetic) permeability $\mu$.
For a realistic physical model of a uniaxial medium we refer to~\cite{obukhov:2012}, where a relativistic nematic fluid is discussed.

General linear dielectric permeable media are defined as media whose constitutive law, in terms of the dependence of the electric excitation vector $\mathcal{D}$ and the magnetic induction vector $\mathcal{H}$ on the electric and magnetic field vectors $E$ and $B$, is such that~\cite{Perlick}
\begin{equation}\label{eq:permmed}
  \mathcal{D}_a = \epsilon^b{}_a E_b,
  \quad
  \mathcal{H}_a = \mu^b{}_a B_b.
\end{equation}
This form of the constitutive law can easily be translated in the more general covariant framework of pre-metric electrodynamics which we used throughout this article.
Introducing a Lorentzian spacetime metric $g$ and a reference observer with unit time direction $U$, \ie, $g(U,U)=-1$, we identify the electric and magnetic field with respect to the observer from the field strength tensor~$F$
\begin{equation*}
  E_a = F_{ab} U^b,
  \quad
  B_a = -\frac{1}{2} \abs{g}^\frac12 \varepsilon_{abcd} U^b F^{cd}.
\end{equation*}
The corresponding electric excitation and the magnetic induction are obtained from the induction tensor $H$ via
\begin{equation*}
  \mathcal{D}_a = \frac12 \abs{g}^\frac12 \varepsilon_{abcd} U^b H^{cd},
  \quad
  \mathcal{H}_a = -H_{ab} U^b.
\end{equation*}
These definitions of $E,B,\mathcal{D}$ and $\mathcal{H}$ differ from the ones in~\cite{Perlick} due to a different definition of the excitation.
The definitions used here are such that for $\epsilon^a{}_b = \delta^a_b = \mu^a{}_b$ we recover Maxwell electrodynamics.
Combining these equations with the constitutive law~\eqref{eq:permmed}, we see that for consistency the dielectricity and the permeability have to satisfy $\epsilon^a{}_b U^b = 0$ and $\mu^a{}_b U^b = 0$.
Comparing~\eqref{eq:permmed} with the constitutive law of pre-metric electrodynamics~\eqref{eq:conlawcoord}, we see that we can express the constitutive tensor $\kappa_{ab}{}^{cd}$ in terms of the matrices $\epsilon$ and~$\mu$:
\begin{equation*}
  \kappa_{ab}{}^{cd} = 2 \abs{g}^\frac12 \big(\varepsilon_{abfg} \epsilon^{[c}{}_e U^{d]} U^f g^{eg} - \varepsilon_{efgh} \mu^e{}_{[a} U_{b]} U^f g^{cg} g^{dh} \big).
\end{equation*}
Inserting the trivial permeability $\mu^a{}_b = \delta^a_b + U^a g_{bc} U^c$ and the dielectricity $\epsilon^a{}_b = \delta^a_b + U^a g_{bc} U^c + X^a g_{bc} X^c$, where $X$ is a spacelike vector field orthogonal to $U$ on spacetime, we obtain for the constitutive density~\eqref{eq:condens}
\begin{equation}\label{eq:condensuniapp}
  \chi^{abcd} = \frac12 \varepsilon^{abef} \kappa_{ef}{}^{cd} = \abs{g}^\frac12 \big( 2 g^{c[a} g^{b]d} + 4 X^{[a} U^{b]} X^{[d} U^{c]} \big).
\end{equation}
Observe that for vanishing vector field $X$ this constitutive density reduces to the one of Maxwell electrodynamics~\eqref{eq:constmax}.
We thus see that $X$ characterizes the properties of the crystal, in particular its optical axis.

The Fresnel polynomial of the constitutive law under consideration turns out to be bi-metric, \ie, the product of two quadratic polynomials in the wave vectors $k$, each defined through a metric:
\begin{align*}
  \mathcal{G}(k) &= \frac{1}{4!} \varepsilon_{c_1 a_1 a_2 a_3} \varepsilon_{d_3 b_1 b_2 b_3} \chi^{a_1 c_1 b_1 d_1} \chi^{a_2 c_2 b_2 d_2} \chi^{a_3 c_3 b_3 d_3} k_{d_1} k_{c_2} k_{d_2} k_{c_3} \\
  &= \abs{g}^\frac12 g^{-1}(k,k) \big( g^{-1}(k,k) - U(k)^2 g(X,X) + X(k)^2 \big).
\end{align*}
We remark that uniaxial crystals are not the only media with bi-metric Fresnel polynomials but that two other classes of constitutive laws with this property exist~\cite{dahl:2013}.
The matrix $\mathcal{Q}_{ab}$ takes the form
\begin{align*}
  \mathcal{Q}_{ab} = \tilde{\mathcal{Q}}_{ab} - k_{(a} \big( X_{b)} X(k) - U_{b)} U(k) g(X,X) \big) - k_a k_b,
\end{align*}
where we extracted the term
\begin{align*}
  \tilde{\mathcal{Q}}_{ab}
  &= g_{ab} \big( g^{-1}(k,k) + X(k)^2 - U(k)^2 g(X,X) \big) \\
  &\mathrel{\phantom{=}}{} + \big( X(k) U_a - U(k) X_a \big) \big( X(k) U_b - U(k) X_b \big).
\end{align*}
that is not proportional to $k_a$ or $k_b$ for its importance in different gauge choices.
Note that one could use $\tilde{\mathcal{Q}}_{ab}$ to define Green's operators in a generalization of Feynman gauge to uniaxial crystals, see also~\eqref{eq:inv_feynman}.
Due to the rich structure of the Fresnel polynomial we can use both kinds of gauge choices for $\kappa$ discussed in Sect.~\ref{sub:gauge}.
This leads to the following gauge equivalent quasi-inverses:
\begin{enumerate}
  \item Since the Fresnel polynomial is the product of two metrics, we can use the first factor $g^{ab}$ to construct
  \begin{equation*}
    \kappa^a_1 = \frac{g^{ab} k_b}{g^{-1}(k,k)}.
  \end{equation*}
  \item Equally well we could use the second factor $g^{ab} - U^a U^b g(X,X) + X^a X^b$ to find
  \begin{equation*}
    \kappa^a_2 = \frac{g^{ab} k_b -U (k) g(X,X) U^a + X(k) X^a}{g^{-1}(k,k) - U(k)^2 g(X,X) + X(k)^2}.
  \end{equation*}
  \item The canonical choice, purely determined by the Fresnel polynomial and thus applicable in any case of pre-metric electrodynamics is
  \begin{equation*}
    \kappa^a_3 = \frac{\mathcal{G}^{abcd} k_b k_c k_d}{\mathcal{G}(k)} = \frac12 (\kappa^a_1 + \kappa^a_2).
  \end{equation*}
\end{enumerate}
A quasi-inverse of the principal symbols for field equations of the electromagnetic field inside the uniaxial crystal is now obtained by combining the objects we displayed here explicitly as derived in~\eqref{eq:qinv3form}.

All choices of $\kappa$ yield a different quasi-inverse.
Following Sects.~\ref{sub:inv} and~\ref{sub:causp}, each of these can be used to construct solutions to the field equations by an application to a conserved current.
Clearly the resulting (different) vector potentials are gauge equivalent.
We would like to stress that from the viewpoint of pre-metric electrodynamics the third gauge choice is the most natural one.
The other choices rely on the fact that there is at least one metric in the constitutive law with which a gauge fixing can be defined.
For an uniaxial crystal with constant constitutive law~\eqref{eq:condensuniapp}, the gauge conditions $\kappa$ and the corresponding quasi-inverses of the principal symbol of the field equations can now be used to construct the inverse of the field equations~\eqref{eq:inv}, the Pauli--Jordan propagator~\eqref{eq:pjprop} and the symplectic space of solutions as done in Sect.~\ref{sub:sympl}.

\section{Partial differential operators and the propagation of singularities}\label{app:propsing}

In Sect~\ref{sub:quasiinv} we derived the quasi-inverse of the principal symbol of the field equations.
Here we would like to comment on the propagation of singularities of the theory which are interpreted as the propagation of light rays in the geometric optical limit of pre-metric electrodynamics.

The theory of partial differential equations tells us that \emph{the singularities of the solutions of a partial differential equation $P A = J$ propagate along the flow of the Hamilton vector field $X_P$} associated to the operator's principal symbol $M(x, k)$ if the operator is of \emph{real principal type}, \cf~\cite{dencker:1982}.

An $n \times n$ ($n$ equations for $n$ variables) partial differential operator $P$ is of real principal type if and only if, in addition to its principal symbol $\mathcal{M}$, there exists another $n \times n$ symbol $\mathcal{N}$ such that
\begin{equation}\label{eq:rpt}
  \mathcal{N} \circ \mathcal{M} = \mathcal{S}(x,k) \id,
\end{equation}
where $\mathcal{S}(x, k)$ is a scalar symbol of real principal type.
The associated Hamilton vector field is given by
\begin{equation}\label{eq:hamilton_vector}
  X_P(x, k) \defn \frac{\partial \mathcal{S}(x, k)}{\partial k_a} \frac{\partial}{\partial x^a} - \frac{\partial \mathcal{S}(x, k)}{\partial x^a} \frac{\partial}{\partial k_a},
\end{equation}
and $\mathcal{S}$ is of real principal type if $X_P$ is not radial and not vanishing where $\mathcal{S}(x, k) = 0$.

To apply this concept to pre-metric electrodynamics one must be careful.
It is not obvious that the field equations are an $n \times n$ system of coupled partial differential equations due to the gauge freedom.
However, as sketched in the next paragraph, the field equations of pre-metric electrodynamics are a $3 \times 3$ system of real principal type under certain conditions.

With the introduction of $\mathcal{M}$ in~\eqref{eq:MtoM} by combining $M$ with an $\varepsilon$ tensor density we can rephrase the real principal type condition~\eqref{eq:rpt} for pre-metric electrodynamics as
\begin{equation*}
  \mathcal{S} \id = \mathcal{N} \circ \mathcal{M} = N \circ M.
\end{equation*}
From~\eqref{eq:MQ} we identify $\mathcal{S}$ with the Fresnel polynomial $\mathcal{G}$ and $\mathcal{N}$ with the bi-linear $\mathcal{Q}$.
To indeed obtain an identity instead of a projector on the right hand side of~\eqref{eq:MQ} we need to restrict to the gauge-fixed subspace $V$.
We conclude that the principal symbol $M$ of pre-metric electrodynamics is a $3 \times 3$ symbol of real principal type if the scalar symbol~$\mathcal{G}$ is of real principal type.
The singularities of the solutions of the theory propagate along the integral curves of the Hamilton vector field determined by the zeroes of the Fresnel polynomial.
Unfortunately, as we see in the example of Maxwell electrodynamics~\eqref{eq:maxwell_fresnel}, $\mathcal{G}$ is not necessarily of real principal type.
In this case we can, however, divide both $Q$ and $\mathcal{G}$ by $g^{-1}(k,k)$ to find that $\mathcal{M}$ is still of real principal type.
This is in fact an example of a general strategy that one should apply in such a situation: If $\mathcal{G}$ is reducible, reduce it.

\small
\bibliographystyle{cmphref}
\bibliography{GLED.bib}

\end{document}